\documentclass{ieeeaccess}
\usepackage{cite}
\usepackage{amsmath,amssymb,amsfonts}
\usepackage{graphicx}
\usepackage{textcomp}
\usepackage{booktabs}
\usepackage{listings}
\usepackage{xcolor}
\let\accessyear\year
\def\year{2026}
\usepackage{tikz}
\usetikzlibrary{arrows.meta, positioning, shapes.geometric, fit, backgrounds, calc}
\let\year\accessyear
\year{2026}% the class's own macro: sets the footer year
\definecolor{accessblue}{cmyk}{1,0.3,0,0.2}
\makeatletter
\long\def\@makecaption#1#2{%
  \ifx\@captype\@IEEEtablestring
    \begin{flushleft}
    \vspace*{5pt}
    {\vss\color{accessblue}\tablecapheadfont #1. \ }{\raggedright\tablecapfont#2\vss}%
    \end{flushleft}
    \@IEEEtablecaptionsepspace
  \else
    \@IEEEfigurecaptionsepspace
    {\raggedright\noindent{\color{accessblue}\figcapheadfont #1. \ }\figcapfont#2\par}%
  \fi}
\makeatother
\makeatletter\input{t1ptm.fd}\input{t1phv.fd}\input{t1pcr.fd}\makeatother
\DeclareFontShape{T1}{ptm}{n}{n}{<->ssub*ptm/m/n}{}
\DeclareFontShape{T1}{ptm}{n}{it}{<->ssub*ptm/m/it}{}
\DeclareFontShape{T1}{ptm}{n}{sl}{<->ssub*ptm/m/sl}{}
\DeclareFontShape{T1}{ptm}{n}{sc}{<->ssub*ptm/m/sc}{}
\DeclareFontShape{T1}{phv}{n}{n}{<->ssub*phv/m/n}{}
\DeclareFontShape{T1}{phv}{n}{it}{<->ssub*phv/m/it}{}
\DeclareFontShape{T1}{pcr}{n}{n}{<->ssub*pcr/m/n}{}
\makeatletter
\def\ps@titlepage{%
  \def\@oddhead{\parbox[t]{\textwidth}{\mbox{}\\[-6.3mm]\mbox{}\nheaderfont\color{accessblue}\headName\hfill\mbox{}\color{black}\vspace*{-1.5mm}\par\hrulefill}}
  \def\@evenhead{\parbox[t]{\textwidth}{\mbox{}\\[-6.3mm]\mbox{}\headerlogo\hfill\nheaderfont\color{accessblue}\headName{\raisebox{-2pt}{\,}}\color{black}\vspace*{-1.5mm}\par\hrulefill}}
  \def\@oddfoot{\raisebox{3pt}{%
    \hbox to 0pc{\hbox to \textwidth{{\footervolfont VOLUME\ \thevol, \theyear}\hfill\mbox{}}}%
    \hbox to 0pc{\hbox to \textwidth{\mbox{}\hfill{\footervolfont\@pubid}\hfill\mbox{}}}%
    \hbox to 0pc{\hbox to \textwidth{\mbox{}\hfill{\footerpagefont\thepage}}}}}%
  \def\@evenfoot{\raisebox{3pt}{%
    \hbox to 0pc{\hbox to \textwidth{{\footerpagefont\thepage}\hfill}\mbox{}}\hspace*{1.8pc}%
    \hspace*{-1.8pc}\hbox to 0pc{\hbox to \textwidth{\mbox{}\hfill{\footervolfont\@pubid}\hfill\mbox{}}}%
    \hbox to 0pc{\hbox to \textwidth{\mbox{}\hfill{\footervolfont VOLUME\ \thevol, \theyear}}}\hspace*{1.8pc}}}}
\def\ps@headings{%
  \def\@oddhead{\parbox[t]{\textwidth}{\mbox{}\\[-6.3mm]{\raisebox{-2pt}{\headerfont\rightmark}}\hfill\mbox{}\vspace*{-1.5mm}\par\hrulefill}}
  \def\@evenhead{\parbox[t]{\textwidth}{\mbox{}\\[-6.3mm]\mbox{}\headerlogoall\hfill{\raisebox{-2pt}{\headerfont\leftmark}}\vspace*{-1.5mm}\par\hrulefill}}
  \def\@oddfoot{\raisebox{8pt}{%
    \hbox to 0pc{\hbox to \textwidth{{\footervolfont VOLUME\ \thevol, \theyear}\hfill\mbox{}}}%
    \hbox to 0pc{\hbox to \textwidth{\mbox{}\hfill{\footerpagefont\thepage}}}}}%
  \def\@evenfoot{\raisebox{8pt}{%
    \hbox to 0pc{\hbox to \textwidth{{\footerpagefont\thepage}\mbox{}\hfill}}\hspace*{1.8pc}%
    \hspace*{-1.8pc}\hbox to 0pc{\hbox to \textwidth{\mbox{}\hfill{\footervolfont VOLUME\ \thevol, \theyear}}}\hspace*{1.8pc}}}}
\ps@headings
\makeatother
\usepackage{etoolbox}
\makeatletter
\patchcmd{\appendices}{\def\thesectiondis{APPENDIX \Alph{section}}.}{\def\thesectiondis{APPENDIX \Alph{section}}}{}{\ClassWarning{ieeeaccess}{appendices patch failed}}
\makeatother
\makeatletter

\newif\if@biographyTOCentrynotmade \@biographyTOCentrynotmadetrue
\makeatother
\usepackage[hyphens]{url}
\usepackage[hidelinks,bookmarks=false]{hyperref}
\usepackage{multirow}
\newcommand{\Gswe}{G_{SWE}}
\newcommand{\Gsrc}{G_{SRC}}
\newcommand{\Gpr}{G_{PR}}
\newcommand{\Gex}{G_{PR}^{exist}}
\newcommand{\Gnew}{G_{PR}^{new}}
\newcommand{\finding}[1]{\vspace{2pt}\noindent\fbox{\parbox{0.96\columnwidth}{\small\textit{#1}}}\vspace{2pt}}

\makeatletter
\newcommand{\val}[1]{\@ifundefined{val@#1}{\textbf{??#1}}{\csname val@#1\endcsname}}
\makeatother
\InputIfFileExists{numbers}{}{}
\providecommand{\texorpdfstring}[2]{#1}
\DeclareRobustCommand{\aicite}{\cite{claudecode}}
\begin{document}
\history{Date of publication xxxx 00, 0000, date of current version xxxx 00, 0000.}
\doi{10.1109/ACCESS.2026.DOI}

\title{Exploration Structure in LLM Agents for Multi-File Change Localization}
\author{\uppercase{Akeela Darryl Fattha}\authorrefmark{1},
\uppercase{Kia Ying Chua}\authorrefmark{1},
\uppercase{Lingxiao Jiang}\authorrefmark{1},
and \uppercase{Laura Wynter}\authorrefmark{1}}
\address[1]{School of Computing and Information Systems, Singapore Management University, 80 Stamford Road, Singapore 178902 (e-mail: akeelaf.2022@scis.smu.edu.sg; kiayingchua.2023@scis.smu.edu.sg; lxjiang@smu.edu.sg; lwynter@smu.edu.sg)}
\tfootnote{This work was supported by the Singapore Ministry of Education (MOE) Academic Research Fund (AcRF) Tier 1 Grant under Proposal ID 24-SIS-SMU-074.}

\markboth
{Fattha \headeretal: Exploration Structure in LLM Agents for Multi-File Change Localization}
{Fattha \headeretal: Exploration Structure in LLM Agents for Multi-File Change Localization}

\corresp{Corresponding author: Laura Wynter (e-mail: lwynter@smu.edu.sg).}

\begin{abstract}
Fixing an issue starts with finding the files to change, and most benchmark issues touch several. LLM agents for this task increasingly add structure, such as specialized tools, multiple agents, or code summaries. We ask which of this structure matters when the model is small. We build a coordinator agent with repository tools that limit their output and a reusable registry of subsystem-scoped domain agents. We compare six configurations, each adding structure to the previous one. They are the model alone, a single agent with a Python REPL, a coordinator with only the REPL, the previous structure plus repository tools, then plus the registry map, and finally plus consultation of the domain agents. We also run four published localizers with the same model, two larger-model references, and five non-LLM rankers on 105 multi-file issues from three Python repositories, in audited single-commit checkouts. Most of the gain comes from letting the model explore the repository over many turns. Pooled source-file F1 rises from 0.199 for the model alone to 0.677 for the REPL-only coordinator. The repository tools add a small, non-significant gain, and the registry does not help, lowering pooled F1 by 0.031. A small model in a plain multi-turn agent loop (0.695) thus exceeds all four published localizers (best 0.586), while Codex with a larger model reaches 0.751. For a small model, the agent loop matters more than added agents. New files and narrative documentation remain largely missed. We release the benchmark, predictions, tool logs, and scripts.
\end{abstract}

\begin{keywords}
Bug localization, change impact analysis, file localization, large language models, LLM agents, multi-agent systems, SWE-bench
\end{keywords}

\titlepgskip=-15pt

\maketitle

\section{Introduction~\texorpdfstring{\protect\aicite}{}}
\label{sec:intro}

\PARstart{F}{ixing} a software issue starts with deciding \emph{where} to change the code. This step, file-level change localization, has been studied for more than a decade as information-retrieval (IR) based bug localization~\cite{zhou2012buglocator,saha2013bluir,ye2014l2r,lee2018bench4bl}, and it is now the first stage of most large language model (LLM) issue-resolution pipelines~\cite{xia2024agentless,zhang2024autocoderover,yu2025orcaloca,chen2025locagent}. Localization errors propagate, because a repair stage cannot edit a file that was never retrieved.

The task has become harder. In SWE-bench Verified~\cite{swebbenchverified}, 14\% of the 500 issues modify more than one file. In SWE-bench Pro~\cite{swebbenchpro}, 86\% of the 731 public instances do (Section~\ref{sec:benchmark}). A typical multi-file fix touches a command-line entry point, the library it calls, and sometimes the user documentation. Finding all of these files requires visiting several regions of the repository.

Most LLM agents explore a repository with a single agent that issues one action per turn (list a directory, read a file, run a search) and conditions the next action on the result~\cite{yang2024sweagent,wang2024openhands,zhang2024autocoderover,kang2024autofl}. Prior work documents two problems with this sequential exploration. First, the agent's context accumulates everything it has read, and LLM accuracy degrades as context grows and as irrelevant material is added~\cite{liu2024lostmiddle,shi2023distracted}. Second, trajectory studies of software engineering agents report that unsuccessful runs spend many steps on unproductive navigation and repeated actions~\cite{bouzenia2025trajectories}. Repository structure makes these problems concrete. In the Ansible repository studied here, \texttt{test/} contains about six times as many files as \texttt{lib/} (3{,}270 versus 532 at the first 2020 base commit), so an agent that enters the test tree reads many plausible but irrelevant files. A further error concerns predicted paths that do not exist in the checked-out revision. Some of them are plausible new files, such as a test module or a changelog fragment that a fix might add. Others are source paths that the model recalls from other versions of the project or invents. Methods without repository access produce many of the latter (Table~\ref{tab:newfiles}).

An alternative is to give the agent structure. The repository can be split into coherent subsystems, each assigned to a specialist agent with its own scoped context, and a coordinator can decide which specialists to consult. Multi-agent LLM systems for software engineering exist~\cite{he2025llmmas,hong2023metagpt,tao2024magis,chen2024coder}, but they divide work by \emph{role} (planner, developer, tester) and not by \emph{repository region}. It has not been measured whether region-based decomposition helps multi-file localization, or which part of such a design is responsible for any improvement.

This paper is an empirical study of that question. Repository tools that return a limited amount of text per call, such as a file viewer that shows part of a file or a search that returns a bounded number of matches, are established. SWE-agent~\cite{yang2024sweagent} introduced them as part of an agent--computer interface. Our design reuses such tools in a coordinator agent and adds one specific component, a reusable registry of subsystem-scoped domain agents that the coordinator may consult. Each part can be enabled separately. Holding the model fixed (Claude Haiku 4.5), we compare a sequence of six configurations. They are a plain LLM without repository access, a single-agent Recursive Language Model (RLM)~\cite{rlmpaper} that explores through a Python read-eval-print loop (REPL), the coordinator with the REPL as its only tool, the coordinator with the repository tools, the coordinator that sees the map of the registry but cannot consult it, and the full system. Most of the gain comes from letting the model explore the repository over many turns. Pooled source-file F1 rises from 0.199 without repository access to 0.677 for the coordinator with only a REPL. The repository tools add a small gain that is not significant, because without them the model writes equivalent code in the REPL, and the registry does not help. A small model in a plain multi-turn agent loop therefore exceeds four published localizers run with the same model, while larger models remain more accurate. With the same model we also run four published localization methods, Agentless~\cite{xia2024agentless}, CoSIL~\cite{jiang2025cosil}, LocAgent~\cite{chen2025locagent}, and Reformulate-Retrieve-Localize (RRL)~\cite{caumartin2026rrl}. We add two larger-model references, Claude Sonnet 4.6 and the Codex command-line interface (CLI), and five non-LLM rankers. The rankers are BM25, a vector space model (VSM), the revised vector space model (rVSM) of BugLocator, a ranker that uses only file paths, and a learning-to-rank model in the style of Ye et al.~\cite{ye2014l2r}. Every agentic run works in a checkout that contains only the issue's base commit, under targeted guard checks on network access, \texttt{git}, and file paths, and an audit of every logged output checks what each agent saw.

We study 105 multi-file issues from the three Python repositories of SWE-bench Pro with the most instances. These are Ansible, in three time windows (2020, 2025, and 2026), Open Library, and qutebrowser. We address five research questions (RQs).
\begin{itemize}
\item RQ1 (Effectiveness). How accurately do the methods localize the files changed by a fix, compared with published localizers and larger models?
\item RQ2 (Attribution). Which components of the exploration structure explain the differences when the model is held fixed?
\item RQ3 (Guiding the coordinator). Does it help to instruct the coordinator to consult the domain agents for every issue, or to give it the candidate files of a cheaper localizer?
\item RQ4 (Systematic misses). Which kinds of changed files do all methods miss?
\item RQ5 (Robustness). Do the conclusions hold across repositories and across time, including issues filed after the models' training data were collected?
\end{itemize}

The paper makes four contributions.
\begin{enumerate}
\item The finding that, for a small model, the multi-turn agent loop accounts for most of the accuracy and exceeds four published localizers run with the same model, while output-limited tools and a subsystem registry add no measurable gain, and that newly created files and narrative documentation are the files all methods miss (Sections~\ref{sec:results} and~\ref{sec:discussion}).
\item A study of exploration structure for multi-file localization with the model held fixed, in a sequence of configurations that adds repository access, the agent harness, repository tools that limit their output, the map of a subsystem registry, and consultation of subsystem-scoped domain agents. (Sections~\ref{sec:approach} and~\ref{sec:design}).
\item A comparison with four published LLM localizers run with the same model, scored both on their full output and at a cutoff chosen on held-out windows, with two larger-model references and five non-LLM rankers, on 105 issues from three repositories and five time windows. The analysis reports run-level and issue-level uncertainty, paired confidence intervals for differences, multiplicity-corrected tests, a Python-only evaluation, and an evaluation without the issue texts written for this study (Section~\ref{sec:results}).
\item An evaluation protocol with single-commit checkouts, targeted guard checks on the network access, programs, and file paths of agent code, and an audit of every logged action and output of all runs, a list of evaluation pitfalls that can distort comparisons of agents, and a replication package with the benchmark, all predictions and tool logs, the registries, and the scripts that regenerate every table and figure (Section~\ref{sec:protocol} and the replication package).
\end{enumerate}

%% ---------------------------------------------------------------------------
\section{Background and Related Work~\texorpdfstring{\protect\aicite}{}}
\label{sec:related}

\subsection{IR-Based and Learning-to-Rank Bug Localization}
IR-based bug localization ranks source files by their textual similarity to a bug report. BugLocator~\cite{zhou2012buglocator} introduced the revised vector space model (rVSM), which multiplies cosine similarity by a document-length prior and adds similar fixed reports. BLUiR~\cite{saha2013bluir} indexes code structure (classes, methods, identifiers) separately, AmaLgam~\cite{wang2014amalgam} adds version history and similar reports, BRTracer~\cite{wong2014brtracer} segments files and uses stack traces, BLIA~\cite{youm2015blia} combines reports, stack traces, and change history, and Locus~\cite{wen2016locus} ranks change hunks. Topic models have also been used~\cite{nguyen2011bugscout,lukins2010bugloc}. Comparative studies show that simple models are strong baselines~\cite{rao2011retrieval} and that results depend heavily on how the benchmark is built~\cite{lee2018bench4bl,akbar2020irbugloc}.

Learning-to-rank (L2R) combines such signals. Ye et al.~\cite{ye2014l2r,ye2016mapping} train a ranking support vector machine (SVM) on features that include lexical similarity, similarity enriched with application programming interface (API) documentation, collaborative filtering over past reports, class-name matches, and the recency and frequency of past fixes. Deep models learn the relation between reports and code directly. HyLoc and DNNLOC combine neural relevance with rVSM~\cite{lam2015hyloc,lam2017dnnloc}, NP-CNN learns joint features from natural and programming language~\cite{huo2016npcnn}, DeepLocator uses an enhanced convolutional neural network (CNN)~\cite{xiao2017deeplocator}, and RLocator formulates ranking as reinforcement learning~\cite{chakraborty2024rlocator}. Surveys cover this line of work~\cite{sharma2024mlcode-survey,zhang2023codeloc-survey,hou2024llm4se}. These methods return a ranked list, so a cutoff $K$ is needed to compare them with methods that return a set. We therefore report both rank metrics and set-based F1 at a $K$ chosen on held-out data (Section~\ref{sec:baselines}).

\subsection{LLM Agents for Issue Resolution and Localization}
SWE-bench~\cite{jimenez2024swebench} and its successors~\cite{swebbenchverified,swebbenchpro} made end-to-end issue resolution the standard LLM evaluation for software engineering. Agentless~\cite{xia2024agentless} runs a fixed pipeline of hierarchical localization (file, then function, then line) followed by repair. SWE-agent~\cite{yang2024sweagent} and OpenHands~\cite{wang2024openhands} give an LLM an agent--computer interface with shell, editor, and search tools. AutoCodeRover~\cite{zhang2024autocoderover} adds search based on the abstract syntax tree (AST), SpecRover~\cite{ruan2025specrover} extends it with summaries of the intended behavior of functions that guide its search and review, SWE-Search~\cite{antoniades2025swesearch} adds Monte Carlo tree search (MCTS) over patches, and Lingma SWE-GPT~\cite{ma2024lingmaswegpt} trains a model on development-process data. Several systems target localization specifically. AutoFL~\cite{kang2024autofl} lets an LLM navigate code with function calls to explain faults, RepoGraph~\cite{ouyang2025repograph} and CoSIL~\cite{jiang2025cosil} navigate code graphs, LocAgent~\cite{chen2025locagent} builds a heterogeneous graph of the repository for agent-guided search, OrcaLoca~\cite{yu2025orcaloca} schedules actions by priority and prunes context, SweRank~\cite{reddy2026swerank} applies retrieve-and-rerank to issue localization, and RRL~\cite{caumartin2026rrl} lets an agent reformulate the bug report into BM25 queries and inspect the retrieved files. Al Awad and Ivanov~\cite{alawad2026costeffective} compare cheaper and more expensive explorer models for localization and find that cheaper explorers keep 73\% to 92\% of the F1 of the reference with 84\% to 95\% fewer tokens. MAsFL~\cite{pham2026masfl} also targets small models, with several small language models acting as cooperating agents for fault localization. RepoUnderstander~\cite{ma2024repounderstand} studies how agents build repository-level context. Empirical studies compare these agents and analyze their trajectories and patches~\cite{meng2026agentsbugfix,bouzenia2025trajectories,chen2025evaluatingagents}.

\subsection{Agent--Computer Interfaces}
\label{sec:aci}
SWE-agent~\cite{yang2024sweagent} showed that the interface between an LLM agent and the repository affects how well the agent works. Its file viewer shows a window of 100 lines that the agent scrolls, its search commands list at most 50 results and ask for a narrower query otherwise, and its ablations found the 100-line window more effective than both a smaller window and the full file. The repository tools of our coordinator follow the same principle (Section~\ref{sec:tools}). They return a short file in full and only the beginning of a long one, at most 50 search matches, and a compact directory tree. We therefore do not claim these tools as a contribution. What our study adds is a measurement of their effect on multi-file localization in which the agent, its prompt, and its step limit are held fixed and only the tools change, and a reusable registry of subsystem-scoped agents on top of them.

\subsection{How Agent Systems Decompose the Work}
Table~\ref{tab:axes} compares the decomposition axis of representative systems. Single-agent systems improve the granularity of each step but keep one sequential trajectory. SWE-Search branches over alternative patches, not over repository regions. MetaGPT~\cite{hong2023metagpt}, MAGIS~\cite{tao2024magis}, CodeR~\cite{chen2024coder}, and HyperAgent~\cite{phan2024hyperagent} run several agents, but each agent has a \emph{role} (manager, repository custodian, developer, tester, navigator) and sees the whole repository. A recent survey of LLM multi-agent systems for software engineering finds that role decomposition is the dominant design~\cite{he2025llmmas}. Our domain agents are instead specialists for a \emph{region} of the repository (for example \texttt{lib/ansible/galaxy/}). They persist across issues and are consulted only when the coordinator considers their region relevant. The two axes are complementary.

\begin{table}[t]
\centering
\caption{Decomposition axis of representative LLM systems for issue localization and resolution.}
\label{tab:axes}
\scriptsize
\setlength{\tabcolsep}{3pt}
\begin{tabular}{p{2.6cm}p{2.0cm}p{3.0cm}}
\toprule
System & Agents & Decomposition axis \\
\midrule
SWE-agent, OpenHands, AutoCodeRover, AutoFL & one & none (sequential steps) \\
Agentless & pipeline & stage (file $\to$ function $\to$ line) \\
RRL & one & query reformulation + BM25 \\
LocAgent, CoSIL, RepoGraph & one & code graph navigation \\
OrcaLoca & one & prioritized action queue \\
SWE-Search & one per branch & alternative patches (MCTS) \\
MAGIS, CodeR, HyperAgent, MetaGPT & several & task role \\
RLM~\cite{rlmpaper} & recursive calls & sub-problems of a long input \\
\textbf{This work} & coordinator + registry & \textbf{repository subsystem} \\
\bottomrule
\end{tabular}
\end{table}

\subsection{Recursive Language Models}
Recursive Language Models (RLMs)~\cite{rlmpaper} keep a long input outside the LLM context, in the variables of a persistent Python REPL, and let the model write code to inspect, filter, and summarize it, optionally calling sub-LLMs on parts of it. RLMs improve long-context aggregation over single long documents~\cite{rlmpaper,longbenchv2}. Applied to a repository, an RLM can call \texttt{os.listdir}, \texttt{grep}, and \texttt{open} from the REPL, and we use this configuration as our single-agent baseline. Our coordinator keeps the RLM idea that bulky content should stay in a programmatic environment and adds tools that are designed for repositories and limit how much text they return (Section~\ref{sec:approach}). We compare both forms of access with the same model (Section~\ref{sec:rq2}).

\subsection{Logical Coupling}
Files that are not statically coupled often change together~\cite{gall1998logicalcoupling,zimmermann2005change}. Documentation is a typical example, since changes to user-facing behavior require documentation updates that the issue text rarely mentions. RQ4 measures how much of this coupling LLM-based localizers recover.

%% ---------------------------------------------------------------------------
\section{Subsystem-Scoped Domain Agents~\texorpdfstring{\protect\aicite}{}}
\label{sec:approach}

Figure~\ref{fig:arch} shows the design. It has three components, each of which can be disabled for ablation.

\begin{figure*}[t]
\centering
\begin{tikzpicture}[
  font=\footnotesize, x=1cm, y=1cm,
  box/.style={draw, rounded corners=2pt, align=center, minimum height=8mm, inner sep=3pt, fill=white},
  store/.style={draw, rounded corners=2pt, align=center, inner sep=3pt, fill=blue!6},
  agent/.style={draw, rounded corners=2pt, align=center, minimum height=6mm, minimum width=24mm, inner sep=2pt, fill=green!8},
  arr/.style={-{Stealth[length=2mm]}, semithick},
  lab/.style={font=\scriptsize, fill=white, inner sep=1pt}]
% Phase 1: initialization
\node[box] (repo) at (0,0) {Repository at\\base commit};
\node[box] (coord1) at (3.6,0) {Coordinator\\(light exploration)};
\node[box] (builder) at (7.6,0) {Agent builder\\(scope validation)};
\node[store] (reg) at (12.3,0) {Agent registry: 8 to 45 domain agents\\(map: name, folders, purpose;\\per agent: familiarization summary)};
\draw[arr] (repo) -- node[lab, above=1pt]{tree} (coord1);
\draw[arr] (coord1) -- node[lab, above=1pt]{domains} (builder);
\draw[arr] (builder) -- node[lab, above=1pt]{register} (reg);
% Phase 2: per issue
\node[box] (issue) at (0,-3.2) {Issue text};
\node[box, minimum width=27mm] (coord2) at (3.6,-3.2) {Coordinator\\(Haiku 4.5)};
\node[box, fill=orange!8, text width=44mm] (io) at (2.3,-5.0) {\textbf{Repository tools}: \texttt{read\_file} (preview, line ranges), \texttt{search\_codebase} ($\leq$50 matches), \texttt{list\_files} (compact tree), \texttt{parse\_file}, \texttt{python\_repl}};
\node[agent] (a1) at (9.6,-2.45) {\texttt{galaxy}};
\node[agent] (a2) at (9.6,-3.2) {\texttt{cli\_interface}};
\node[agent] (a3) at (9.6,-3.95) {\dots};
\node[font=\scriptsize, align=left, anchor=west] at (11.0,-3.2) {domain agents consulted\\independently, each with\\its scoped summary and\\the same repository tools};
\node[box] (out) at (9.6,-5.25) {Final answer \texttt{\{"files": [...]\}}};
\draw[arr] (issue) -- (coord2);
\draw[arr, <->] (3.0,-3.6) -- ({3.0,-3.6} |- io.north);
\draw[arr, <->] (coord2.east) -- node[lab, above=3pt, pos=0.5, sloped]{question / candidates} (a1.west);
\draw[arr, <->] (coord2.east) -- (a2.west);
\draw[arr, <->, dashed] (coord2.east) -- (a3.west);
\draw[arr] (coord2.south east) -- (4.95,-5.25) -- node[lab, above=1pt, pos=0.6]{merge, filter} (out.west);
\draw[arr, gray!70!black] (reg.south) |- node[lab, pos=0.75, above=1pt]{registry summary} (3.6,-1.6) -- (coord2.north);
\begin{scope}[on background layer]
\node[draw=gray!50, dashed, rounded corners, fit=(repo)(reg), inner sep=6pt, label={[anchor=south west, font=\footnotesize\bfseries]north west:Phase 1: initialization (once per repository)}] {};
\node[draw=gray!50, dashed, rounded corners, fit=(issue)(io)(a1)(a3)(out)(coord2), inner sep=6pt, label={[anchor=north west, font=\footnotesize\bfseries]north west:Phase 2: per issue}] {};
\end{scope}
\end{tikzpicture}
\caption{Architecture. Phase 1 builds a registry of subsystem-scoped domain agents once per repository. In Phase 2 the coordinator explores with the repository tools and may consult relevant domain agents, each of which answers from its own scoped context. The coordinator merges and filters the candidates into the final file set. The \emph{coordinator-only} ablation removes the registry and consultation, the \emph{registry-map} ablation keeps the registry summary in the prompt but removes consultation, the \emph{REPL-only} ablation also replaces the repository tools by \texttt{python\_repl}, the \emph{nudged} variant instructs the coordinator to consult the agents for every issue, and the \emph{seeded} variant adds the candidate files of a cheaper localizer to the coordinator prompt.}
\label{fig:arch}
\end{figure*}

\subsection{Repository Tools}
\label{sec:tools}
The coordinator reads the repository through tools that limit how much text one call returns. The file-reading tool, \texttt{read\_file}, returns a file of up to 300 lines and 80{,}000 characters in full. For a longer file it returns only the first 50 lines and the total length, together with an instruction to request other line ranges or to search within the file, and the model then requests only the parts it needs. The search tool, \texttt{search\_codebase}, runs ripgrep and returns at most 50 matching lines with their paths and line numbers. The listing tool, \texttt{list\_files}, shows a directory as an indented tree instead of one full path per line. The structure tool, \texttt{parse\_file}, lists the classes and functions of a file. \texttt{python\_repl} provides a persistent Python environment in which intermediate results can be kept in variables, so that code, and not the LLM context, filters them. Inside it, the model can call the other tools and an \texttt{llm(prompt)} helper that sends a prompt to the same model. Its output is capped at 100{,}000 characters per call. The harness also registers two file-editing tools, \texttt{write\_file} and \texttt{update\_file}. The task prompt forbids edits. In one issue run a coordinator wrote its own answer to a new file in its checkout, which is rebuilt before every issue, and no run edited a repository file. When the registry is enabled, \texttt{consult\_agents} is added (Section~\ref{sec:registry}).

The coordinator runs as a LangGraph state graph with a limit of 100 steps per issue, in which a model turn and the execution of its tool calls are separate steps, so an issue allows about 49 tool-calling turns. When the limit is reached, the coordinator receives one more turn, without tools, in which it is asked to answer from what it has found. Domain agents run in their own graphs with the same limit. Section~\ref{sec:protocol} reports how often the limit was reached.

\subsection{Coordinator}
The coordinator is an LLM agent that receives the task prompt (replication package, \texttt{prompts.md}), the issue text, and, when the registry is enabled, a summary of the registered domain agents. It explores with the repository tools and returns a JavaScript Object Notation (JSON) list of files with a short rationale. With the registry disabled, the result is the \emph{coordinator-only} configuration, a single agent with the repository tools.

\subsection{Registry of Domain Agents}
\label{sec:registry}
In an initialization phase, run once per repository and base commit, the coordinator briefly explores the repository and proposes maintenance areas, each given by a name and a set of folders. A \emph{maintenance area} is a set of directories that implements one user-visible capability, such as \texttt{lib/ansible/galaxy/} for the \texttt{ansible-galaxy} tool. An agent-builder step validates the proposed scopes against the repository tree and registers one domain agent per area. Each domain agent reads its region and writes a familiarization summary with its purpose, responsibilities, and key files. The registry therefore holds two kinds of information. The coordinator's prompt contains a \emph{map} with one line per domain agent, giving its name, its folders, and a one-sentence description of what it covers. The familiarization summaries stay with the domain agents and reach the coordinator only through consultation. The registries built in our runs contain 8 to 45 domain agents for Ansible, 15 to 30 for Open Library, and 10 to 18 for qutebrowser. Every run builds its own registry at the earliest base commit of its window and reuses it for all issues of that run. The state of the coordinator and of the domain agents is restored to the end of initialization before each issue, so that no information passes from one issue to the next. Initialization costs 0.37M to 1.73M tokens, which is amortized over all queries and reported separately.

At query time the coordinator may call \texttt{consult\_agents(question, paths)}. The matching domain agents answer independently, using the same repository tools with file access restricted to their assigned folders, and return candidate files with explanations. Agents do not see each other's context, and several agents can be consulted in one call. In practice the coordinator consulted at least one domain agent in 31\% to 67\% of issues, depending on the repository (Section~\ref{sec:rq3}). Domain agents therefore act mainly as subsystem experts consulted on demand and rarely as a large parallel search.

\subsection{Guiding the Coordinator}
\label{sec:guiding}
We study two ways of guiding the coordinator, each against the default \emph{adaptive} policy, which tells the coordinator to consult agents ``if available folder agents clearly match the relevant subsystem''. The \emph{nudged} policy replaces this sentence with an instruction to route every problem to the relevant agents first and to use the coordinator's own tools afterwards to verify their input (replication package, \texttt{prompts.md}). The \emph{seeded} variant keeps the adaptive policy and adds the candidate files of a cheaper localizer to the prompt. Before the coordinator starts exploring, its prompt lists the files that the file stage of Agentless~\cite{xia2024agentless} returned for the issue, together with the instruction to use them as starting points, to verify them, and to look for other files that need edits. The candidates cost about 4k tokens per issue. Each variant changes only the prompt of the default system and uses the same harness, step limit, timeout, and protocol. It builds its own registry in the same way, so the variants and the default system also differ in their registries.

%% ---------------------------------------------------------------------------
\section{Study Design~\texorpdfstring{\protect\aicite}{}}
\label{sec:design}

\subsection{Research Questions and Motivation}
\label{sec:rqs}

\textit{RQ1 (Effectiveness). How accurately do the methods localize the files changed by a fix?}
Before attributing effects to components, we need to know whether structured exploration with a small model is competitive with established non-LLM rankers, with published LLM localizers that use the same model, with larger models, and with a commercial coding agent. We report precision, recall, micro-F1, and the number of issues whose gold set is fully recovered.

\textit{RQ2 (Attribution). Which components explain the differences when the model is held fixed?}
The methods differ in several ways at once. They differ in whether they can read the repository, in the agent harness that runs the model, in how much text each access returns, in whether a registry exists, and in how many tokens they use. We therefore compare a sequence of configurations with the model held fixed. The plain LLM has no repository access. The RLM explores through a Python REPL in its own runner, so the first step changes repository access together with the prompt, the runner, and the number of model calls. The coordinator with the REPL as its only tool runs in the coordinator's harness with the same prompt and step limit as the full system. Comparing it with the RLM therefore measures the harness as a whole, including the prompt, the tool-calling interface, the turn budget, and the timeout. Comparing it with the coordinator that has the repository tools changes only the tools, which are added to the REPL, and this is the one comparison in the sequence that isolates a single component. The coordinator that sees the registry map but cannot consult it, and the full system, then compare configurations that differ in the information that the registry provides. We also report token use for every configuration and an ensemble baseline that uses extra inference without exploring.

\textit{RQ3 (Guiding the coordinator). Does it help to instruct the coordinator to consult the domain agents for every issue, or to give it the candidates of a cheaper localizer?}
The coordinator decides on its own where to look and whom to consult. Two kinds of guidance could improve this decision. If domain agents help, consulting them for every issue might help further, or it might add candidates from unrelated subsystems. A cheap localizer could point the coordinator to likely files at the start. This could reduce the exploration needed, but it could also bias the search toward the errors of the localizer. We run each variant once on all 105 issues, because of cost, and compare it with the default system. We also test whether combining the final outputs of the default system and of Agentless helps without changing the coordinator.

\textit{RQ4 (Systematic misses). Which kinds of changed files do all methods miss?}
A localizer that finds the code but never the documentation, or never the files that a fix creates, fails in a way that average F1 does not show. We analyze documentation files and newly created files separately.

\textit{RQ5 (Robustness). Do the conclusions hold across repositories and time?}
A design built around subsystems may suit one architecture better than another, models may have seen older issues and their fixes during pre-training, and repository structure changes over time. We therefore repeat the evaluation on two further repositories with different architectures and on Ansible windows from 2020, 2025, and 2026.

\subsection{Benchmark Construction}
\label{sec:benchmark}

\paragraph*{Repository selection} We draw from SWE-bench Pro~\cite{swebbenchpro} because 85.6\% of its 731 public (v1) instances modify at least two files, compared with 14.2\% for SWE-bench Verified (both computed from the gold patches). We applied two criteria. The first (C1) is that the repository is written in Python, so that all methods, including the lexical tokenizer and the RLM REPL, operate on the same language. The second (C2) is that it has at least 70 SWE-bench Pro instances, so that a dense six-month window exists. Exactly three repositories satisfy both criteria (Table~\ref{tab:repos}), and we use all three. They differ in architecture. Ansible is an automation tool with a plugin architecture and many separable subsystems, Open Library is a web application, and qutebrowser is a desktop web browser. Ansible has the highest share of multi-file instances (88\%) and of documentation co-changes, so we study it in three time windows. Section~\ref{sec:ttv} discusses the remaining threats to generality.

\begin{table}[t]
\centering
\caption{The SWE-bench Pro (v1) Python repositories, all of which are used. Multi is the share of instances whose gold patch modifies at least two files, and Docs is the share that touches documentation files.}
\label{tab:repos}
\scriptsize
\begin{tabular}{lrrrr}
\toprule
Repository & Instances & Multi & Files/inst. & Docs \\
\midrule
ansible/ansible & 96 & 0.88 & 3.80 & 0.18 \\
internetarchive/openlibrary & 91 & 0.80 & 3.58 & 0.02 \\
qutebrowser/qutebrowser & 79 & 0.71 & 2.84 & 0.00 \\
\bottomrule
\end{tabular}
\end{table}

\paragraph*{Time windows} We use windows of at most six months (183 days), so that the repository structure is stable within a window. For 2020 we moved a 183-day window over the dates of all 96 Ansible instances and selected the window with the most instances. This is May 15 to November 9, 2020, with 19 instances by issue date, and no non-overlapping 183-day window contains more than 15. For RQ5 we built two additional Ansible windows, April to September 2025 and January to May 2026. SWE-bench Pro contains no Ansible instances from that period. Anthropic documents a training-data cutoff of July 2025 for Claude Haiku 4.5 (\texttt{claude-haiku-4-5-20251001}) and of January 2026 for Claude Sonnet 4.6 (\texttt{claude-sonnet-4-6})~\cite{anthropicmodels}. All 19 issues of the 2026 window are dated after the Haiku 4.5 cutoff, and 16 of them after the Sonnet 4.6 cutoff. The 2025 window overlaps the training period of both models, with 13 of its 19 issues dated before August 2025. The Codex model has no published cutoff (Section~\ref{sec:baselines}). Following the SWE-bench Pro instance format, each instance is a GitHub issue resolved by a merged pull request (PR) that modifies at least one non-test file under \texttt{lib/}. The candidates were merged PRs labeled \texttt{bug} or \texttt{feature} that change at least one runtime source file and only add or modify files. Low-signal maintenance changes (typos, lint, type hints) and large mechanical sweeps were excluded. Within each window, 19 PRs were first sampled with date stratification within each difficulty level, so that the issues span the whole period with a difficulty mix close to that of the 2020 window. The authors then reviewed this sample by hand and replaced weak tasks with PRs from the same window of similar size and difficulty, preferring PRs with clearer user-visible behavior, with issue, discussion, or test evidence for a self-contained task, and with more subsystem diversity, and excluding tasks that could not be stated without describing the patch. The rule is documented with the selection code, but the individual replacements were not logged, so we cannot list which PRs were replaced.

Many of these PRs have no self-contained issue text, so the issue texts were written by a Codex CLI worker, one per instance. The worker received the PR number, title, and URL, the linked issues, and the number of changed files. It was instructed to gather the rest itself with the GitHub CLI and web search, namely the PR description, comments, commits, and list of changed files, the linked issue discussions, and the full PR diff. It was told to rely on the issue and its discussion first and to use the diff only as evidence of user-visible behavior. The texts were to describe the problem as a GitHub issue would and were not to contain internal source paths, line numbers, descriptions of the patch, private helper names, or the implementation strategy. Exact public names that users see, such as command-line options, configuration keys, module and plugin names, public classes and functions, and error messages, were allowed. The worker ran without a sandbox, and its model and raw outputs were not retained. The only automatic check was a schema check of the output, so we checked the texts afterwards. None of the 108 $\Gsrc$ files of the two windows is named by path or file name. Four of the 19 Ansible 2025 texts contain an underscore-prefixed name, three of them environment or configuration keys and one taken from an error trace. To measure how much the texts reveal about the fix, we extracted the functions, classes, modules, and configuration keys that each resolving change defines or modifies and counted how many of them the issue text names, with the same procedure for all 105 issues. The generated texts name a median of 0.10 of these identifiers, compared with 0.10 for the SWE-bench Pro texts (Mann--Whitney $p=0.92$). For identifiers that cannot be ordinary words, such as names with underscores or in camel case, the shares are 0.05 and 0.11 on average ($p=0.17$). The generated texts name no changed file, whereas 7 of the 67 SWE-bench Pro texts do. They name the stem of a changed source file, typically a user-facing module or plugin name such as \texttt{galaxy} or \texttt{winrm}, in 13 of 38 issues, compared with 18 of 67 (Fisher $p=0.51$). We therefore find no evidence that the generated texts reveal more about the fix than the SWE-bench Pro texts, but they were written with access to the fix, and Section~\ref{sec:rq5} reports all results also without these 38 issues. The selection and generation code and the instruction are in the experiment harness repository, and the overlap analysis is in the replication package.

For Open Library and qutebrowser we applied the 2020 rule directly to their SWE-bench Pro instances and used every instance in the 183-day window with the most instances by fix-commit date. This gives 22 Open Library issues (July 2023 to January 2024) and 26 qutebrowser issues (November 2020 to April 2021). Their issue texts are the SWE-bench Pro problem statements, as for Ansible 2020. Table~\ref{tab:benchmark} summarizes the five windows.

\paragraph*{Base commits} Every instance is evaluated at its own base commit (the parent of the fix), as in SWE-bench Pro. The domain-agent registry of a run is built at the earliest base commit of the window and reused for all of its issues.

\begin{table*}[t]
\centering
\caption{Benchmark windows. An issue is hard if $|G_{SRC}|\geq 2$. Commits is the number of distinct base commits, since each issue is evaluated at its own base commit. $G_{PR}^{exist}$ and $G_{PR}^{new}$ split $G_{PR}$ by whether the file exists at the base commit. Docs counts the issues whose $G_{SRC}$ contains a documentation file under \texttt{docs/} or \texttt{doc/}. The 2025 and 2026 Ansible base commits contain no documentation directory. $G_{SWE}$ counts the files of the SWE-bench Pro gold patches, and the 2025 and 2026 Ansible issues are not in SWE-bench Pro.}
\label{tab:benchmark}
\small
\setlength{\tabcolsep}{4pt}
\resizebox{\textwidth}{!}{%
\begin{tabular}{llrrrrrrrrr}
\toprule
Repository & Issue dates & Issues & Hard & Commits & $|G_{SRC}|$ & $|G_{PR}|$ & $|G_{PR}^{exist}|$ & $|G_{PR}^{new}|$ & Docs & $|G_{SWE}|$ \\
\midrule
\texttt{ansible/ansible} & 2020-05 -- 2020-11 & 19 & 16 & 19 & 66 & 171 & 120 & 51 & 10 & 63 \\
\texttt{ansible/ansible} & 2025-04 -- 2025-09 & 19 & 14 & 19 & 64 & 197 & 122 & 75 & 0 & -- \\
\texttt{ansible/ansible} & 2026-01 -- 2026-05 & 19 & 13 & 19 & 44 & 121 & 81 & 40 & 0 & -- \\
\texttt{internetarchive/openlibrary} & 2023-07 -- 2024-01 & 22 & 21 & 22 & 80 & 119 & 109 & 10 & 0 & 81 \\
\texttt{qutebrowser/qutebrowser} & 2020-11 -- 2021-04 & 26 & 19 & 26 & 70 & 119 & 110 & 9 & 10 & 80 \\
\midrule
Total & & 105 & 83 &  & 324 & 727 & 542 & 185 & 20 &  \\
\bottomrule
\end{tabular}}
\end{table*}

\subsection{Gold Sets}
\label{sec:gold}

A localizer's score depends on which files count as required. We define four gold sets for each issue $i$.
\begin{itemize}
\item $\Gpr$ contains every file changed by the resolving PR, including tests and changelog fragments. For SWE-bench Pro instances these are the files changed between the base commit and the fix commit named in the instance identifier.
\item $\Gsrc$, the primary gold set, contains the files of $\Gpr$ that are not tests, changelogs, or build, continuous integration (CI), and developer-tooling metadata. Tests are files under a \texttt{test}, \texttt{tests}, or \texttt{testing} directory and files named \texttt{test\_*.py} or \texttt{conftest.py}. Changelogs are files under \texttt{changelogs/} or named \texttt{changelog*}. Metadata comprises \texttt{.github/}, \texttt{.azure-pipelines/}, \texttt{packaging/}, \texttt{hacking/}, \texttt{scripts/dev/}, \texttt{licenses/}, and requirement and setup files. The same rule is applied in every repository and window, and documentation remains in $\Gsrc$.
\item $\Gex$ and $\Gnew$ contain the files of $\Gpr$ that do and do not exist at the base commit, respectively. A file in $\Gnew$ is created by the fix. Predicting its exact path is a different task from finding an existing file, so we score it separately.
\item $\Gswe$ contains the files of the SWE-bench Pro gold patch and is available for all windows except Ansible 2025 and 2026. For Ansible 2020 it is a strict subset of $\Gsrc$ (63 versus 66 files). We report it for comparability with SWE-bench Pro.
\end{itemize}
We use $\Gsrc$ as the primary gold set because it is defined by a reproducible rule that applies to all windows, whereas $\Gswe$ is not available for the 2025 and 2026 Ansible issues. An issue is \emph{hard} if $|G_{SRC,i}|\geq 2$.

\subsection{Methods Compared}
\label{sec:baselines}

Table~\ref{tab:configs} lists the configurations. All LLM methods receive the same task prompt (replication package, \texttt{prompts.md}).

\begin{table*}[t]
\centering
\caption{Configurations. Runs gives the number of independent repetitions per window (Ansible 2020/2025/2026, Open Library, qutebrowser). Every agentic configuration works in a per-issue checkout that contains only the base commit, under the access controls of Section~\ref{sec:protocol}. The coordinator and domain-agent configurations share one harness, prompt, step limit (100 graph steps, about 49 tool-calling turns, then one final turn without tools), and per-issue timeout (600\,s). Rows marked $\star$ form the sequence of configurations of RQ2. The Sonnet RLM and the two RQ3 variants were run once per window because of cost.}
\label{tab:configs}
\scriptsize
\setlength{\tabcolsep}{3pt}
\begin{tabular}{p{3.4cm}p{2.2cm}p{5.2cm}p{3.2cm}c}
\toprule
Configuration & Model & Repository access & Budget & Runs \\
\midrule
$\star$ Plain LLM & Haiku 4.5 & none (single call) & one call & 5/3/3/3/3 \\
Plain LLM & Sonnet 4.6 & none (single call) & one call & 3/3/3/3/3 \\
$\star$ Single-agent RLM & Haiku 4.5 & Python REPL in the RLM runner, one REPL action per turn & $\leq$30 root turns, 3{,}600\,s & 3/3/3/3/3 \\
Single-agent RLM & Sonnet 4.6 & as above & as above & 1/1/1/1/1 \\
$\star$ Coordinator, REPL only & Haiku 4.5 & \texttt{python\_repl} only, coordinator harness & step limit, 600\,s & 3/3/3/3/3 \\
$\star$ Coordinator only & Haiku 4.5 & repository tools (Section~\ref{sec:tools}) & step limit, 600\,s & 3/3/3/3/3 \\
$\star$ Coordinator + registry map & Haiku 4.5 & as above, registry map in the prompt, no consultation & step limit, 600\,s & 3/3/3/3/3 \\
$\star$ Domain agents, adaptive & Haiku 4.5 & as above, with consultation (default system) & step limit per agent, 600\,s & 3/3/3/3/3 \\
Domain agents, nudged & Haiku 4.5 & as adaptive, consultation instructed for every issue & as above & 1/1/1/1/1 \\
Domain agents, seeded & Haiku 4.5 & as adaptive, Agentless candidates in the prompt & as above & 1/1/1/1/1 \\
Codex CLI & GPT-6.1-Terra, high reasoning & shell in a read-only sandbox without network & 600\,s & 3/3/3/3/3 \\
Agentless, CoSIL (file stage) & Haiku 4.5 & repository structure, Python files & default pipeline & 3/3/3/3/3 \\
LocAgent & Haiku 4.5 & code graph and BM25 tools, Python files & default (2 samples, merged) & 3/3/3/3/3 \\
RRL (reimplemented) & Haiku 4.5 & BM25 search, file view (512 tokens) & $\leq$25 turns & 3/3/3/3/3 \\
BM25, VSM, rVSM, path BM25, L2R & none & index of the base commit & top-$K$ & deterministic \\
\bottomrule
\end{tabular}
\end{table*}

\paragraph*{Plain LLM} The plain LLM receives the issue text in a single API call, without repository access. It measures what the model knows about the project from pre-training. We run Claude Haiku 4.5, the model of the agentic systems, and Claude Sonnet 4.6 as a larger reference.

\paragraph*{Single-agent RLM} This is our application of the RLM approach~\cite{rlmpaper}. The model works in a persistent Python REPL seeded with the repository path, from which it may list directories, search with grep, and open files, and it submits its answer with \texttt{FINAL([...])} or by setting an answer variable. The model writes each REPL action as text, and the runner executes it and shows the output in the next turn. The runner ends each assistant message after its first complete action, so that the model sees the output of every action before it writes the next one, and it does not accept an answer in the first turn, before any output has been seen. The iteration cap counts root turns. The Haiku RLM used a median of 24 turns per issue and the Sonnet RLM 21.

\paragraph*{Codex CLI} Codex is a commercial coding agent with its own model. We use \texttt{gpt-6.1-terra}, the default model of the CLI (version 0.160.0) on the run date, with high reasoning effort. Codex runs a shell in its read-only sandbox, which blocks network access and writes but not reads outside the checkout, in the same per-issue checkout as the other methods and with the same prompt. Its full event log is audited with the other logs (Section~\ref{sec:protocol}). Its token counts are those reported by the CLI. It is not part of the ablation but a reference point for a widely used agent with a larger model that explores freely.

\paragraph*{Published LLM localizers} We run four recent localization methods with Claude Haiku 4.5, the model of our agentic systems, so that differences are not due to the model. Each sees the issue text and the repository at the issue's base commit, and each is run three times. \emph{Agentless}~\cite{xia2024agentless} localizes hierarchically. We run its file-level stage, in which the model reads the repository structure and the issue and returns a ranked list of up to five files. \emph{CoSIL}~\cite{jiang2025cosil} also starts from the repository structure and then searches a module call graph that it builds during the search, with reflection steps that prune candidates. We use its file-level stage. \emph{LocAgent}~\cite{chen2025locagent} indexes the repository as a heterogeneous graph of directories, files, classes, and functions, together with a BM25 index, and an agent searches it with graph tools. We use its default configuration with function calling, two samples, and its merging of the sampled file rankings. \emph{RRL}~\cite{caumartin2026rrl} has no public implementation, so we reimplemented it from the paper. The model first extracts a structured description of the issue. Its explanation, identifiers, and code snippets form a BM25 query ($k_1{=}0.9$, $b{=}0.4$) over the non-test Python files. An agent may then run further BM25 searches (top 20) and view the first 512 tokens of a file. It returns five files and checks them once in a self-validation step. Agentless, CoSIL, and RRL use temperature 0 as in their original settings, so their three runs differ little. All four methods consider only Python files, as in their original implementations, and therefore cannot predict documentation or other non-Python files, which make up 18\% of the $\Gsrc$ files. We therefore also report every method on the Python files of $\Gsrc$ alone. The four methods return ranked lists, which we score in two ways, as the complete list that the method returns and at a cutoff $K$ chosen on held-out windows as for the non-LLM rankers (Section~\ref{sec:metrics}). We changed the tools only where needed to run them on our instances. We load instances from local files and extract each base commit from a local clone, add the Haiku model identifier and price, and enable prompt caching for RRL and LocAgent, which changes cost but not outputs. Haiku often returned malformed tool arguments to LocAgent, for example lists encoded as strings. Without a correction LocAgent stopped on many issues, so we added a step that repairs these arguments and we catch the resulting parse errors. LocAgent did not terminate on one qutebrowser issue in one run, which counts as an empty prediction. RRL exceeded the context window on three issues per run, and in these cases we keep its last submitted answer. The replication package lists every change.

\paragraph*{Non-LLM rankers} To cover the classical and learning-to-rank localization literature, we implement five rankers. Each indexes the text files at the issue's own base commit, namely Python, reStructuredText, AsciiDoc, Markdown, plain text, YAML, HTML, SQL, Less, CSS, JavaScript, JSON, TOML, INI, shell, PowerShell, C\#, and Jinja files, which cover every file type of $\Gsrc$. In the PR gold sets, 31 of 727 files, such as extensionless test fixtures and \texttt{.stdout} files, are not indexed. The rankers are BM25~\cite{robertson2009bm25} over path and content tokens ($k_1{=}1.2$, $b{=}0.75$), VSM with the cosine similarity of term frequency and inverse document frequency (TF-IDF) vectors, the length-weighted rVSM of BugLocator~\cite{zhou2012buglocator}, BM25 over path tokens only, and an L2R model in the style of Ye et al.~\cite{ye2014l2r}. The L2R model is a pairwise linear ranking SVM over the four scores above, file-name mentions in the issue, path-token overlap, the recency and frequency of changes in the preceding year (taken from the Git history), and file length. On Ansible it is trained leave-one-window-out, with two windows for training and the third for testing. For Open Library and qutebrowser it is trained on all Ansible windows, so these two repositories form a cross-project test. Because the rankers return lists, we report micro-F1 at the cutoff $K$ that maximizes F1 on the training windows, so that $K$ is never tuned on the evaluated data. We also report mean average precision (MAP).

\subsection{Metrics and Statistical Analysis}
\label{sec:metrics}
For predictions $\hat F_i$ and gold set $G_i$, precision, recall, and F1 are
\begin{equation}
P=\frac{\sum_i|\hat F_i\cap G_i|}{\sum_i|\hat F_i|},\;
R=\frac{\sum_i|\hat F_i\cap G_i|}{\sum_i|G_i|},\;
F_1=\frac{2PR}{P+R}.
\end{equation}
We use micro-averaging because issues have between 1 and 11 gold files and micro-F1 weights every gold file equally. Under $\Gsrc$ and $\Gswe$ we remove predicted files that the $\Gsrc$ rule classifies as tests, changelogs, or metadata before scoring, for every method. A method is then not penalized for also listing the tests it would update, and methods that consider only source files are compared on equal terms. The rankers' candidate lists are filtered in the same way before the top-$K$ cutoff. Under $\Gpr$, $\Gex$, and $\Gnew$, all predictions are scored. An issue whose gold set is empty under a definition is left out of that definition, which leaves 104 issues under $\Gex$, because one Open Library fix only creates files. \emph{All-gold} (AG) counts the issues with $G_i\subseteq\hat F_i$. For $\Gnew$ we report the recall of exact paths. For every method we also count the predicted paths that neither exist at the base commit nor are created by the fix, separately for source paths and for proposed tests, changelogs, and other non-source files.

The four published localizers and the non-LLM rankers return ranked lists. We score the published localizers both on the complete list they return and on its first $K$ files, where $K$ is chosen, separately for each window, as the value that maximizes pooled micro-F1 under $\Gsrc$ on the other four windows. Out-of-scope files are removed before the cutoff, as for the rankers. The non-LLM rankers are scored only at a held-out $K$.

LLM runs vary, so we report two kinds of uncertainty. Run-to-run variation is given by the 95\% Student-$t$ interval over independent runs. Variation due to the sample of issues, which dominates with 19 to 26 issues per window, is given by a 95\% percentile interval from 5{,}000 bootstrap resamples of issues, in which all runs of a resampled issue are kept together. For results pooled over the 105 issues, each issue's counts are averaged over runs, so that windows with five runs do not dominate.

We use two summaries that answer different questions and keep them apart. \emph{Pooled micro-F1} sums true positives, false positives, and false negatives over all issues and so weights issues by their number of gold and predicted files. \emph{Mean per-issue F1} averages each issue's F1 and weights every issue equally. For a comparison of two methods we report the difference in pooled micro-F1 and the mean per-issue difference, each with a 95\% paired bootstrap interval in which the same resampled issues are used for both methods. We test the per-issue differences with the two-sided Wilcoxon signed-rank test~\cite{wilcoxon1945} on the 105 paired issues. The test does not assume normality and is appropriate for small samples~\cite{arcuri2011practical}. We fixed a family of 21 planned comparisons, adjust the $p$-values with the Holm procedure, and report the matched-pairs rank-biserial correlation $r$ as the effect size. Cliff's $\delta$~\cite{cliff1993dominance,romano2006cliffdelta} leads to the same conclusions and is included in the replication package. A difference that is not significant is reported as such, together with its interval, and not as evidence that the two methods are equivalent.

\subsection{Evaluation Protocol and Access Controls}
\label{sec:protocol}
Every method that reads the repository works at the base commit of each issue, and the non-LLM rankers index it. The plain LLM uses no repository. A checkout at the base commit contains the right source files, but a clone of the repository also contains its later history, including the resolving commit, and the issue page and the fix are on the web. We therefore run every agentic configuration under three controls.
\begin{itemize}
\item \emph{Single-commit checkouts.} Before each issue, the agent's working directory is rebuilt from the tree of the base commit and committed as the only commit of a fresh repository. The full clone used to export the tree lies outside the working directory, so \texttt{git} inside the checkout shows no later history.
\item \emph{A guard on agent code.} Code that the agent writes, in the coordinator's REPL and in the RLM's REPL, runs under a Python audit hook. It refuses network connections to hosts other than the model API, and refuses to start network clients, \texttt{git}, and further Python interpreters. It also refuses to open or list files, and to start programs with path arguments, under the user's home directory outside the checkout and the Python installation. This excludes the full clones, the benchmark files, the gold sets, other checkouts, and earlier results. The guard checks the paths that agent code passes to Python and to the programs it starts. It does not trace the files those programs open themselves, and the audit below covers this. The file tools of the coordinator are restricted to the checkout by construction. Codex runs in its own read-only sandbox, which blocks network access and writes but not reads outside the checkout, so for Codex the audit is the only check. The configurations that consult domain agents ran under a stricter version of the guard, which gives each run a private temporary directory, refuses the shared ones, and also checks the working directory of programs.
\item \emph{Logging.} Every tool call with its arguments, every REPL action with its output, every refused access, and, for Codex, every command with its output are recorded. In the configurations that consult domain agents, an action log records every tool call of the coordinator and of the domain agents, including registry initialization, with its output.
\end{itemize}
The guard refused 159 actions in 141 of the 2{,}205 agentic issue runs. These were 69 \texttt{git} commands, such as \texttt{git log --all} with a search term from the issue, 15 attempts to start Python, 64 uses of the shared temporary directory under the stricter guard, 3 other paths outside the checkout, and 8 URLs or network programs. The agent received an error each time and continued.

After the runs, we audited the logs of all 2{,}205 agentic issue runs (\texttt{access\_audit.py}). The audit pairs each of the 80{,}403 actions taken on the issues (tool calls, REPL blocks, Codex commands), among them 14{,}634 of domain agents, and the 7{,}675 actions of registry initialization with their outputs, and stops with an error when a log, clone, or commit is missing. Without the action log, the coordinator-only and map-only runs keep REPL inputs and outputs in full but only the first 500 characters of a file-tool output (6{,}813 of 10{,}906 and 6{,}405 of 10{,}831 outputs). These tools are confined to the checkout by construction, so their logged paths establish where their output came from. In the actions it flags history and network commands, paths outside the checkout, benchmark or result file names, and use of the shared temporary directory. In the outputs it flags the hash of the resolving commit, every line of at least 30 characters that the resolving commit adds and that occurs nowhere at the base commit, and, for actions in the temporary directory, every path shown. It checks the 904 registry entries that coordinators see against the resolving changes of every issue of the window. The audit found one weakness. The guard of the configurations without domain agents admitted the shared temporary directory, and in 44 of their issue runs agent code ran \texttt{grep} or \texttt{find} in \texttt{/tmp}, usually after guessing it as the location of the repository. In 4 of these, the output included files of our own analysis tools that listed gold files and answers. These issue runs were repeated on November 19, 2026, with the same registry, under the stricter guard, and the repetitions are reported. Dropping the affected trials altogether changes the pooled F1 of the coordinator-only and map-only configurations by less than 0.001. In the other runs, the temporary directory showed only failed guesses such as \texttt{/tmp/ansible} and an empty file, and the 6 outputs there that name a gold file repeat a path that the agent itself had supplied in a failed access or an existence check. Under the stricter guard, every attempt to use the shared temporary directory, in 52 issue runs, was refused. Except for the temporary directory and one search in the harness directory that returned nothing, every flagged action that would have left the checkout was refused. We observed no retrieval of an answer. No output contains the resolving commit, and no registry entry contains a line that a resolving change adds. Lines added by the fix appear 43 times in 24 issue runs, 41 of them in text the model wrote itself (code the REPL echoes or plans it printed) and 2 in answers of domain agents. The replication package contains this evidence, the coverage per configuration, and a manifest of the repeated runs. The logs record every action of every agent, with the exceptions above for file-tool outputs and for the registry initialization of the map-only runs, whose registry entries were checked instead, and they do not show the files that a program started by an agent opens itself.

Information did not pass between issues. Each issue starts from the state at the end of registry initialization, and coordinator-only runs start from an empty session. Registry initialization sees only the earliest base commit of the window.

The plain LLMs, the published localizers, which read only an index of the base commit, and the non-LLM rankers have no access to later history.

We record how each issue run ended. An issue that reaches the step limit receives a final synthesis turn (Section~\ref{sec:tools}). The step limit was reached in 116 agentic issue runs, among them 30 of 315 of the coordinator-only, 51 of 315 of the REPL-only, and 11 of 315 of the domain-agent issue runs. In all but 1 of them the final turn produced a list of files, and that one, without a parseable list, is scored as empty. 4 issue runs reached the timeout and count as empty, none ended with another error, and an explicitly empty list was returned in 6 Codex issue runs, which stated that the problem was not present at the base commit, and in 1 other issue run. \texttt{outcomes.json} in the replication package classifies how every issue run ended.

%% ---------------------------------------------------------------------------
\section{Results~\texorpdfstring{\protect\aicite}{}}
\label{sec:results}

\begin{table*}[t]
\centering
\caption{Micro-F1 under the source gold set $G_{SRC}$, defined by the same rule in every repository. Each window shows the mean over $n$ independent runs with the 95\% Student-$t$ half-width across runs, omitted when $n=1$. Pooled is the micro-F1 over all 105 issues with a 95\% issue-level bootstrap CI (5{,}000 resamples), which reflects the sampling of issues rather than of runs. The published localizers are scored on their complete ranked lists and, in the rows marked top-$K$, on their first $K$ files with $K$ chosen on the other four windows. Tokens is the mean number of input and output tokens per issue, excluding the one-off registry initialization. The non-LLM rankers return the top-$K$ files. $K$, and the learning-to-rank model, are chosen on other windows only, namely the two other Ansible windows for Ansible and all Ansible windows for the two added repositories. The best value per column is in bold and the second best is underlined.}
\label{tab:main}
\small
\setlength{\tabcolsep}{3.5pt}
\resizebox{\textwidth}{!}{%
\begin{tabular}{lcccccccr}
\toprule
& & \multicolumn{3}{c}{\texttt{ansible}} & & & & \\
\cmidrule(lr){3-5}
Method & $n$ & 2020 & 2025 & 2026 & \texttt{openlibrary} & \texttt{qutebrowser} & Pooled [95\% CI] & Tokens/issue \\
\midrule
Plain LLM (Haiku) & 5/3/3/3/3 & .257\,$\pm$\,.029 & .276\,$\pm$\,.014 & .322\,$\pm$\,.171 & .099\,$\pm$\,.017 & .097\,$\pm$\,.041 & .199 [.152, .253] & 1k \\
Plain LLM (Sonnet) & 3/3/3/3/3 & .465\,$\pm$\,.033 & .322\,$\pm$\,.070 & .497\,$\pm$\,.027 & .223\,$\pm$\,.047 & .372\,$\pm$\,.127 & .362 [.297, .425] & 1k \\
Single-agent RLM (Haiku) & 3/3/3/3/3 & .491\,$\pm$\,.113 & .571\,$\pm$\,.089 & .747\,$\pm$\,.055 & .528\,$\pm$\,.033 & .724\,$\pm$\,.071 & .598 [.538, .657] & 0.32M \\
Single-agent RLM (Sonnet) & 1/1/1/1/1 & .618 & \underline{.698} & .786 & \underline{.653} & \textbf{.831} & \underline{.714 [.660, .767]} & 0.28M \\
\midrule
Coordinator, REPL only (Haiku) & 3/3/3/3/3 & .588\,$\pm$\,.074 & .655\,$\pm$\,.072 & .799\,$\pm$\,.099 & .621\,$\pm$\,.050 & .766\,$\pm$\,.042 & .677 [.628, .724] & 0.65M \\
Coordinator only (Haiku) & 3/3/3/3/3 & .618\,$\pm$\,.072 & .660\,$\pm$\,.173 & .776\,$\pm$\,.015 & .651\,$\pm$\,.037 & .793\,$\pm$\,.034 & .695 [.649, .742] & 0.85M \\
Coordinator + registry map (Haiku) & 3/3/3/3/3 & \underline{.639\,$\pm$\,.019} & .659\,$\pm$\,.016 & \underline{.805\,$\pm$\,.029} & .616\,$\pm$\,.061 & .753\,$\pm$\,.049 & .684 [.634, .734] & 0.85M \\
Domain agents, adaptive (Haiku) & 3/3/3/3/3 & .564\,$\pm$\,.051 & .636\,$\pm$\,.094 & .758\,$\pm$\,.066 & .609\,$\pm$\,.088 & .786\,$\pm$\,.092 & .664 [.614, .715] & 1.12M \\
\midrule
Codex CLI (GPT-6.1-Terra, high) & 3/3/3/3/3 & \textbf{.678\,$\pm$\,.050} & \textbf{.712\,$\pm$\,.031} & \textbf{.872\,$\pm$\,.029} & \textbf{.718\,$\pm$\,.025} & \underline{.811\,$\pm$\,.012} & \textbf{.751 [.701, .800]} & 0.11M \\
\midrule
Reformulate-Retrieve-Localize & 3/3/3/3/3 & .392\,$\pm$\,.000 & .293\,$\pm$\,.000 & .360\,$\pm$\,.000 & .484\,$\pm$\,.000 & .330\,$\pm$\,.015 & .375 [.330, .418] & 0.29M \\
Agentless (file stage) & 3/3/3/3/3 & .444\,$\pm$\,.000 & .352\,$\pm$\,.000 & .475\,$\pm$\,.000 & .355\,$\pm$\,.000 & .373\,$\pm$\,.000 & .395 [.351, .440] & 4k \\
CoSIL (file stage) & 3/3/3/3/3 & .415\,$\pm$\,.000 & .363\,$\pm$\,.018 & .456\,$\pm$\,.000 & .351\,$\pm$\,.012 & .333\,$\pm$\,.014 & .378 [.337, .418] & 4k \\
LocAgent & 3/3/3/3/3 & .550\,$\pm$\,.084 & .475\,$\pm$\,.045 & .614\,$\pm$\,.089 & .578\,$\pm$\,.065 & .499\,$\pm$\,.053 & .540 [.488, .592] & 1.37M \\
\midrule
RRL, held-out top-$K$ & 3/3/3/3/3 & .408\,$\pm$\,.000 & .281\,$\pm$\,.000 & .376\,$\pm$\,.000 & .479\,$\pm$\,.000 & .351\,$\pm$\,.017 & .382 [.337, .427] & 0.29M \\
Agentless, held-out top-$K$ & 3/3/3/3/3 & .525\,$\pm$\,.000 & .345\,$\pm$\,.000 & .562\,$\pm$\,.000 & .372\,$\pm$\,.000 & .382\,$\pm$\,.000 & .429 [.379, .480] & 4k \\
CoSIL, held-out top-$K$ & 3/3/3/3/3 & .423\,$\pm$\,.000 & .399\,$\pm$\,.056 & .504\,$\pm$\,.035 & .371\,$\pm$\,.000 & .377\,$\pm$\,.041 & .409 [.363, .455] & 4k \\
LocAgent, held-out top-$K$ & 3/3/3/3/3 & .589\,$\pm$\,.052 & .466\,$\pm$\,.027 & .705\,$\pm$\,.027 & .627\,$\pm$\,.047 & .560\,$\pm$\,.002 & .586 [.545, .627] & 1.37M \\
\midrule
BM25 & det. & .310 & .314 & .432 & .500 & .448 & -- & -- \\
Learning-to-rank & det. & .390 & .347 & .416 & .476 & .448 & -- & -- \\
\bottomrule
\end{tabular}}
\end{table*}

\subsection{Effectiveness (RQ1)}
\label{sec:rq1}

\begin{figure*}[t]
\centering
\includegraphics[width=\textwidth]{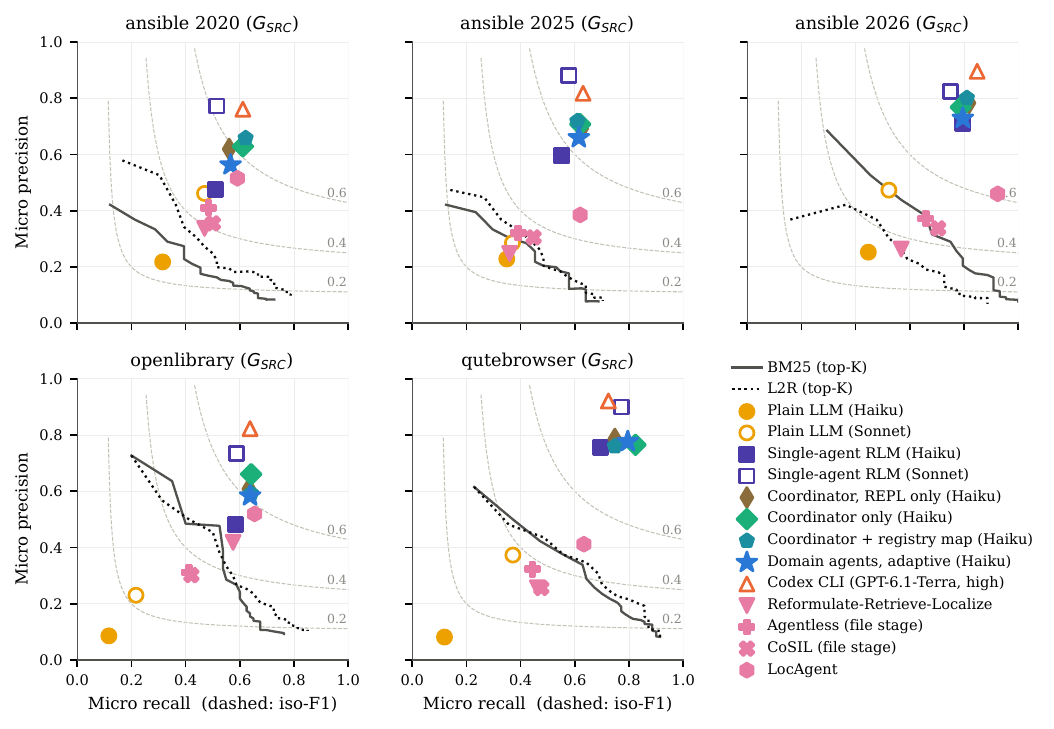}
\caption{Precision and recall under $\Gsrc$ for each window. The lines trace the BM25 and L2R rankers as $K$ grows from 1 to 30, and the markers are the LLM methods (filled for Haiku 4.5, hollow for Sonnet 4.6 and Codex). The four published localizers share one color. Dashed curves are iso-F1 contours at 0.2, 0.4, and 0.6.}
\label{fig:pr}
\end{figure*}

Table~\ref{tab:main} reports micro-F1 under $\Gsrc$ for every window, Table~\ref{tab:allgold} the number of fully recovered issues, and Table~\ref{tab:subsets} the pooled results on the Python files alone and on the issues whose texts come from SWE-bench Pro. Figure~\ref{fig:pr} places each method in precision--recall space together with the BM25 and L2R top-$K$ curves.

Every configuration in which the small model explores the repository over many turns is more accurate than the four published localizers run with the same model, and only the larger models are more accurate still. Pooled over the 105 issues, Codex reaches the highest micro-F1, 0.751 [0.701, 0.800], followed by the Sonnet RLM (0.714). Among the configurations with Haiku 4.5, the coordinator with the repository tools reaches 0.695 [0.649, 0.742], the coordinator with the REPL only 0.677, the coordinator with the registry map 0.684, the full domain-agent system 0.664, and the single-agent RLM 0.598. The Haiku plain LLM reaches 0.199 and the Sonnet plain LLM 0.362. The intervals of the four coordinator configurations overlap almost completely, and Section~\ref{sec:rq2} compares them directly.

All agentic Haiku configurations are well above the four published localizers run with the same model. We make these comparisons with the full domain-agent system, the lowest-scoring of the four coordinator configurations. It is better than RRL, Agentless, and CoSIL by +0.290 to +0.270 in pooled micro-F1 (all $p_{Holm}<0.001$, Table~\ref{tab:tests}). LocAgent is the strongest of the four, with 0.540 on its complete ranked list and 0.586 on its first 3 files, where $K$ is chosen on the other windows. The domain agents remain better than LocAgent at this cutoff, by +0.078 [+0.028, +0.132] pooled and +0.089 [+0.044, +0.135] per issue ($p_{Holm}=0.003$). The published localizers index only Python files. On the Python files of $\Gsrc$, LocAgent reaches 0.582 with its complete list and 0.647 at the cutoff, compared with 0.696 for the domain agents and 0.719 for the coordinator alone, and the difference from the domain agents remains significant ($p_{Holm}=0.016$). LocAgent's complete list has the highest recall of the four localizers (0.812 on Python files), at a precision of 0.453.

Codex is better than the domain agents by $-$0.087 pooled ($p_{Holm}=0.005$), whereas the per-issue difference from the Sonnet RLM, $-$0.017 [$-$0.059, +0.026], is not significant. Both use larger models, and Codex also a different agent design, so these comparisons do not isolate the effect of structure. Under $\Gex$, which also counts tests and other existing files, the order of the methods is the same (Table~\ref{tab:tests}).

\begin{table}[t]
\centering
\caption{All-gold issues under $G_{SRC}$, given as the mean number of issues over runs whose complete source gold set is predicted. The windows contain 19/19/19/22/26 issues. Hard gives the all-gold count on the 83 issues with $|G_{SRC}|\geq 2$.}
\label{tab:allgold}
\resizebox{\columnwidth}{!}{%
\begin{tabular}{lrrrrrrr}
\toprule
Method & Ans.\ 2020 & Ans.\ 2025 & Ans.\ 2026 & OpenLib. & Qutebr. & Total & Hard \\
\midrule
Plain LLM (Haiku) & 2.0 & 3.3 & 6.0 & 1.0 & 1.0 & 13.3 & 5.5 \\
Plain LLM (Sonnet) & 3.0 & 3.3 & 6.0 & 1.0 & 5.0 & 18.3 & 9.7 \\
Single-agent RLM (Haiku) & 5.3 & 6.3 & 13.3 & 7.3 & 13.3 & 45.7 & 26.0 \\
Single-agent RLM (Sonnet) & 5.0 & 7.0 & 11.0 & 6.0 & 15.0 & 44.0 & 24.0 \\
Coordinator, REPL only (Haiku) & 4.7 & 8.0 & 13.0 & 6.7 & 13.7 & 46.0 & 24.7 \\
Coordinator only (Haiku) & 6.3 & 8.0 & 12.7 & 7.7 & 16.7 & 51.3 & 31.7 \\
Coordinator + registry map (Haiku) & 6.3 & 7.7 & 12.7 & 6.7 & 15.0 & 48.3 & 29.0 \\
Domain agents, adaptive (Haiku) & 6.3 & 7.3 & 13.3 & 8.0 & 15.3 & 50.3 & 30.3 \\
Codex CLI (GPT-6.1-Terra, high) & 6.7 & 7.7 & 13.7 & 8.3 & 13.7 & 50.0 & 29.0 \\
Reformulate-Retrieve-Localize & 4.0 & 1.0 & 8.0 & 5.0 & 8.0 & 26.0 & 10.0 \\
Agentless (file stage) & 4.0 & 6.0 & 10.0 & 1.0 & 7.0 & 28.0 & 12.0 \\
CoSIL (file stage) & 4.0 & 8.0 & 10.0 & 1.0 & 9.0 & 32.0 & 14.0 \\
LocAgent & 5.0 & 7.7 & 15.7 & 8.3 & 13.0 & 49.7 & 29.0 \\
\bottomrule
\end{tabular}}
\end{table}

\begin{table}[t]
\centering
\caption{Pooled micro-F1 with 95\% issue-bootstrap CIs on three scopes. $G_{SRC}$ covers all 105 issues. Python only covers the 266 Python files of the 324 $G_{SRC}$ files, with predictions restricted to Python source files, the scope that the published localizers index. Without generated texts covers the 67 issues of Ansible 2020, Open Library, and qutebrowser, whose issue texts come from SWE-bench Pro.}
\label{tab:subsets}
\resizebox{\columnwidth}{!}{%
\begin{tabular}{lccc}
\toprule
Method & $G_{SRC}$ & Python only & Without generated texts \\
\midrule
Plain LLM (Haiku) & .199 [.152, .253] & .220 [.164, .277] & .147 [.097, .202] \\
Plain LLM (Sonnet) & .362 [.297, .425] & .381 [.310, .453] & .347 [.270, .425] \\
Single-agent RLM (Haiku) & .598 [.538, .657] & .649 [.592, .707] & .575 [.509, .644] \\
Single-agent RLM (Sonnet) & .714 [.660, .767] & .754 [.696, .807] & .703 [.644, .761] \\
\midrule
Coordinator, REPL only (Haiku) & .677 [.628, .724] & .714 [.665, .763] & .657 [.606, .708] \\
Coordinator only (Haiku) & .695 [.649, .742] & .719 [.668, .770] & .689 [.640, .738] \\
Coordinator + registry map (Haiku) & .684 [.634, .734] & .721 [.670, .773] & .667 [.610, .722] \\
Domain agents, adaptive (Haiku) & .664 [.614, .715] & .696 [.642, .750] & .653 [.594, .712] \\
\midrule
Codex CLI (GPT-6.1-Terra, high) & .751 [.701, .800] & .785 [.735, .834] & .736 [.680, .789] \\
\midrule
Reformulate-Retrieve-Localize & .375 [.330, .418] & .402 [.357, .446] & .402 [.350, .454] \\
Agentless (file stage) & .395 [.351, .440] & .427 [.379, .474] & .387 [.338, .439] \\
CoSIL (file stage) & .378 [.337, .418] & .406 [.363, .447] & .363 [.313, .413] \\
LocAgent & .540 [.488, .592] & .582 [.527, .634] & .542 [.482, .603] \\
\midrule
RRL, held-out top-$K$ & .382 [.337, .427] & .417 [.370, .465] & .418 [.368, .470] \\
Agentless, held-out top-$K$ & .429 [.379, .480] & .474 [.418, .528] & .421 [.363, .477] \\
CoSIL, held-out top-$K$ & .409 [.363, .455] & .457 [.405, .508] & .400 [.345, .453] \\
LocAgent, held-out top-$K$ & .586 [.545, .627] & .647 [.604, .689] & .584 [.529, .637] \\
\bottomrule
\end{tabular}}
\end{table}

The plain LLMs depend on what the model knows about the project. They do much better on Ansible than on the two other repositories, where they often guess a wrong project layout (for example \texttt{catalog/\dots} instead of \texttt{openlibrary/catalog/\dots}). The non-LLM rankers, with $K$ and the L2R model chosen on other windows, reach 0.347 to 0.432 on Ansible and 0.448 to 0.500 on Open Library and qutebrowser (Table~\ref{tab:ir}). They are below every agentic configuration and above the plain LLMs.

\begin{table*}[t]
\centering
\caption{Non-LLM baselines under $G_{SRC}$, each indexed at the issue's own base commit. F1 is the micro-F1 at the cutoff $K$ selected on other windows, namely the two other Ansible windows for Ansible and all Ansible windows for the added repositories, whose learning-to-rank model is also trained on Ansible only. MAP is the mean average precision of the full ranking. The best F1 per window is in bold.}
\label{tab:ir}
\small
\setlength{\tabcolsep}{4pt}
\begin{tabular}{lrrrrrrrrrr}
\toprule
& \multicolumn{2}{c}{Ans.\ 2020} & \multicolumn{2}{c}{Ans.\ 2025} & \multicolumn{2}{c}{Ans.\ 2026} & \multicolumn{2}{c}{OpenLib.} & \multicolumn{2}{c}{Qutebr.} \\
\cmidrule(lr){2-3}\cmidrule(lr){4-5}\cmidrule(lr){6-7}\cmidrule(lr){8-9}\cmidrule(lr){10-11}
Ranker & F1 & MAP & F1 & MAP & F1 & MAP & F1 & MAP & F1 & MAP \\
\midrule
BM25 & \underline{.310} & 0.35 & \underline{.314} & 0.39 & \textbf{.432} & 0.70 & \textbf{.500} & 0.53 & \textbf{.448} & 0.57 \\
VSM (TF-IDF) & .239 & 0.28 & .302 & 0.33 & .328 & 0.55 & .379 & 0.41 & .360 & 0.46 \\
rVSM (BugLocator) & .296 & 0.31 & .298 & 0.34 & .383 & 0.56 & .405 & 0.47 & \underline{.437} & 0.53 \\
BM25, path only & .113 & 0.23 & .132 & 0.12 & .288 & 0.37 & .137 & 0.17 & .135 & 0.15 \\
Learning-to-rank & \textbf{.390} & 0.48 & \textbf{.347} & 0.41 & \underline{.416} & 0.49 & \underline{.476} & 0.53 & \textbf{.448} & 0.57 \\
\bottomrule
\end{tabular}
\end{table*}

\begin{table}[t]
\centering
\caption{Ansible 2020 window scored against the SWE-bench Pro gold set $G_{SWE}$ (63 files) for comparability with SWE-bench Pro reporting, including the RQ3 variants. HAG is the all-gold count on the 16 issues with $|G_{SRC}|\geq 2$, and $|\hat F|$ is the mean predicted set size. For the non-LLM rankers, $K$ is tuned on this window, which gives an optimistic upper bound.}
\label{tab:swe2020}
\scriptsize
\setlength{\tabcolsep}{2.5pt}
\resizebox{\columnwidth}{!}{%
\begin{tabular}{lrrrrrrr}
\toprule
Method & $n$ & P & R & F1 & AG & HAG & $|\hat F|$ \\
\midrule
Plain LLM (Haiku) & 5 & .218 & .330 & .262\,$\pm$\,.030 & 2.0 & 0.2 & 7.7 \\
Plain LLM (Sonnet) & 3 & .461 & .492 & .476\,$\pm$\,.035 & 4.0 & 2.0 & 3.7 \\
Single-agent RLM (Haiku) & 3 & .476 & .534 & .502\,$\pm$\,.117 & 5.7 & 2.7 & 6.2 \\
Single-agent RLM (Sonnet) & 1 & .773 & .540 & .636 & 5.0 & 2.0 & 3.9 \\
Coordinator, REPL only (Haiku) & 3 & .619 & .587 & .602\,$\pm$\,.076 & 5.0 & 2.0 & 5.3 \\
Coordinator only (Haiku) & 3 & .629 & .640 & .633\,$\pm$\,.076 & 6.7 & 3.7 & 5.9 \\
Coordinator + registry map (Haiku) & 3 & .658 & .651 & \underline{.654\,$\pm$\,.020} & 7.0 & 4.0 & 5.5 \\
Domain agents, adaptive (Haiku) & 3 & .563 & .593 & .577\,$\pm$\,.052 & 7.3 & 4.3 & 6.0 \\
Domain agents, nudged (Haiku) & 1 & .639 & .619 & .629 & 7.0 & 4.0 & 5.7 \\
Domain agents, seeded (Haiku) & 1 & .571 & .635 & .602 & 7.0 & 4.0 & 5.8 \\
Codex CLI (GPT-6.1-Terra, high) & 3 & .761 & .640 & \textbf{.695\,$\pm$\,.050} & 6.7 & 3.7 & 5.6 \\
BM25 (best $K$=5) & det. & .274 & .413 & .329 & 5 & -- & 5 \\
Learning-to-rank (best $K$=3) & det. & .404 & .365 & .383 & 3 & -- & 3 \\
\bottomrule
\end{tabular}}
\end{table}

\finding{Answer to RQ1. With a small model, every configuration that explores the repository over many turns reaches a pooled source-file micro-F1 between 0.598 and 0.695, well above the same model without repository access (0.199) and above four published localizers run with the same model, including LocAgent scored at a held-out cutoff (0.586). Larger models are better, with 0.714 for the Sonnet RLM and 0.751 for Codex.}

\subsection{Contribution of the Components (RQ2)}
\label{sec:rq2}

\begin{table*}[t]
\centering
\caption{Paired comparisons over all 105 issues, all methods with Claude Haiku 4.5 unless marked Sonnet or Codex. $\Delta_{\mu}$ is the difference in pooled micro-F1 and $\Delta_{i}$ the mean difference in per-issue F1 (first minus second method, each issue's F1 averaged over runs), each with a 95\% paired bootstrap interval over issues. W/L counts the issues on which the first method is better or worse. $p_{Holm}$ is from the two-sided Wilcoxon signed-rank test on per-issue F1, Holm-adjusted over the 21 planned comparisons, and $r$ is the matched-pairs rank-biserial correlation. $^\star$Step of the sequence of RQ2. Only ``Repository tools vs REPL only'' changes a single component (Section~\ref{sec:rq2}).}
\label{tab:tests}
\scriptsize
\setlength{\tabcolsep}{3pt}
\begin{tabular}{lccrrrrr}
\toprule
& \multicolumn{5}{c}{$G_{SRC}$} & \multicolumn{2}{c}{$G_{PR}^{exist}$} \\
\cmidrule(lr){2-6}\cmidrule(lr){7-8}
Comparison & $\Delta_{\mu}$ [95\% CI] & $\Delta_{i}$ [95\% CI] & W/L & $p_{Holm}$ & $r$ & $\Delta_{i}$ & $p_{Holm}$ \\
\midrule
Domain agents vs plain LLM & +0.465 [+0.408, +0.525] & +0.483 [+0.421, +0.546] & 93/7 & $<$.001 & +0.98 & +0.430 & $<$.001 \\
Domain agents vs RLM & +0.066 [+0.041, +0.092] & +0.052 [+0.025, +0.078] & 53/22 & .001 & +0.51 & +0.028 & .059 \\
Repository tools vs REPL only$^\star$ & +0.019 [$-$0.003, +0.039] & +0.009 [$-$0.013, +0.031] & 41/27 & 1.000 & +0.14 & +0.003 & 1.000 \\
Coordinator harness vs RLM runner$^\star$ & +0.079 [+0.046, +0.111] & +0.055 [+0.027, +0.084] & 50/25 & .003 & +0.48 & +0.044 & .004 \\
Coordinator only vs RLM & +0.097 [+0.065, +0.130] & +0.064 [+0.035, +0.095] & 50/20 & $<$.001 & +0.56 & +0.048 & .016 \\
Registry (map + consultation)$^\star$ & $-$0.031 [$-$0.052, $-$0.012] & $-$0.013 [$-$0.033, +0.008] & 26/42 & .416 & -0.24 & $-$0.020 & .052 \\
Registry map only$^\star$ & $-$0.011 [$-$0.034, +0.013] & $-$0.004 [$-$0.026, +0.018] & 29/34 & 1.000 & -0.05 & $-$0.005 & 1.000 \\
Consultation, given the map$^\star$ & $-$0.020 [$-$0.046, +0.004] & $-$0.009 [$-$0.030, +0.012] & 32/36 & 1.000 & -0.10 & $-$0.015 & .313 \\
RLM vs plain LLM$^\star$ & +0.399 [+0.338, +0.461] & +0.432 [+0.368, +0.494] & 91/9 & $<$.001 & +0.95 & +0.403 & $<$.001 \\
RLM vs plain LLM (Sonnet) & +0.353 [+0.290, +0.416] & +0.346 [+0.269, +0.423] & 76/14 & $<$.001 & +0.82 & +0.354 & $<$.001 \\
Domain agents vs plain LLM (Sonnet) & +0.303 [+0.239, +0.368] & +0.329 [+0.258, +0.400] & 75/21 & $<$.001 & +0.81 & +0.308 & $<$.001 \\
Domain agents vs RLM (Sonnet) & $-$0.050 [$-$0.085, $-$0.016] & $-$0.017 [$-$0.059, +0.026] & 31/43 & 1.000 & -0.13 & $-$0.046 & .024 \\
Domain agents vs Codex & $-$0.087 [$-$0.124, $-$0.048] & $-$0.060 [$-$0.108, $-$0.012] & 23/52 & .005 & -0.44 & $-$0.047 & .024 \\
Domain agents vs RRL & +0.290 [+0.236, +0.349] & +0.334 [+0.279, +0.389] & 90/15 & $<$.001 & +0.88 & +0.306 & $<$.001 \\
Domain agents vs Agentless & +0.270 [+0.216, +0.322] & +0.282 [+0.227, +0.335] & 87/14 & $<$.001 & +0.87 & +0.270 & $<$.001 \\
Domain agents vs CoSIL & +0.286 [+0.238, +0.339] & +0.324 [+0.272, +0.375] & 93/10 & $<$.001 & +0.93 & +0.290 & $<$.001 \\
Domain agents vs LocAgent & +0.125 [+0.079, +0.170] & +0.132 [+0.084, +0.180] & 67/33 & $<$.001 & +0.54 & +0.154 & $<$.001 \\
Domain agents vs RRL, top-$K$ & +0.283 [+0.226, +0.341] & +0.313 [+0.255, +0.372] & 91/13 & $<$.001 & +0.85 & +0.300 & $<$.001 \\
Domain agents vs Agentless, top-$K$ & +0.235 [+0.175, +0.297] & +0.253 [+0.194, +0.313] & 82/17 & $<$.001 & +0.78 & +0.259 & $<$.001 \\
Domain agents vs CoSIL, top-$K$ & +0.256 [+0.195, +0.318] & +0.273 [+0.213, +0.334] & 85/16 & $<$.001 & +0.81 & +0.284 & $<$.001 \\
Domain agents vs LocAgent, top-$K$ & +0.078 [+0.028, +0.132] & +0.089 [+0.044, +0.135] & 64/37 & .003 & +0.41 & +0.137 & $<$.001 \\
\bottomrule
\end{tabular}
\end{table*}

The Haiku configurations form a sequence (Table~\ref{tab:tests}, rows marked $\star$, and Figure~\ref{fig:forest}). Most of the gain comes from exploration itself, and the structure added on top of it makes little difference. Only the step that adds the repository tools changes a single component. The other steps compare configurations that differ in several respects.

\paragraph*{Repository access} The largest difference lies between the plain LLM and the configurations that explore the repository. The RLM, which explores through a Python REPL, raises pooled F1 from 0.199 for the plain LLM to 0.598 (+0.399 [+0.338, +0.461]), with 91 issues better and 9 worse ($p_{Holm}=<0.001$). The same holds for the larger model, where the Sonnet RLM is better than the Sonnet plain LLM by +0.353. Besides repository access, this step also changes the prompt, the runner, and the number of model calls, so it measures multi-turn exploration as a whole.

\paragraph*{Agent harness} Running the same REPL in the coordinator's harness adds a further gain. With native tool calling and the coordinator's prompt and step limit, pooled F1 rises to 0.677 (+0.079 [+0.046, +0.111], per issue +0.055 [+0.027, +0.084], $p_{Holm}=0.003$). The two configurations differ in how actions are expressed (text parsed by the runner versus native tool calls), in the prompt, and in the turn budget, so this comparison measures the harness as a whole.

\paragraph*{Repository tools} The repository tools make little difference, because without them the model writes the same operations in the REPL. Adding the repository tools to the REPL in the same coordinator changes pooled F1 from 0.677 to 0.695, a difference of +0.019 [$-$0.003, +0.039] pooled and +0.009 [$-$0.013, +0.031] per issue ($p_{Holm}=1.000$). The model used the tools when it had them. The coordinator with the repository tools made 17.0 file reads, 13.7 searches, 3.8 directory listings, and 3.3 REPL calls per issue on average. Without them, it wrote equivalent code in the REPL, typically with \texttt{os.walk}, \texttt{grep}, and \texttt{open}, and reached about the same accuracy. In our setting, therefore, limiting the output of the tools does not explain the accuracy of the coordinator. This step adds the whole tool set, so it does not isolate the output limits, and token use is not held equal. It reached the step limit in 30 of 315 issue runs with the repository tools and in 51 of 315 with the REPL only, and used 0.85M and 0.65M tokens per issue on average.

\paragraph*{Registry} The registry does not help. The full system scores lower than the coordinator alone. With the registry map in its prompt but without consultation, pooled F1 changes by $-$0.011 [$-$0.034, +0.013], and the full system differs from the map-only configuration by a further $-$0.020 [$-$0.046, +0.004]. Together, the full system differs from the coordinator alone by $-$0.031 [$-$0.052, $-$0.012] pooled, an interval that excludes zero, and $-$0.013 [$-$0.033, +0.008] per issue ($p_{Holm}=0.416$, 26 issues better and 42 worse). None of the per-issue differences is significant, and their intervals exclude per-issue improvements larger than 0.018 for the map, 0.012 for consultation given the map, and 0.008 for the full system. The configurations with consultation also ran under the stricter guard, which differs only for actions outside the checkout. Each run built its own registry, so the map-only and consultation runs did not use the same registries, and consultation also adds the computation of the domain agents. Under $\Gex$ the picture is the same. The coordinator consulted at least one domain agent in 49\% of the issues, with 0.72 consultations per issue on average.

\begin{figure}[t]
\centering
\includegraphics[width=\columnwidth]{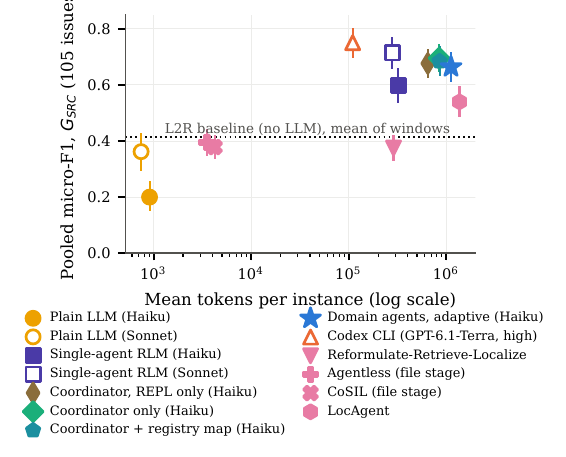}
\caption{Pooled micro-F1 under $\Gsrc$ (105 issues, 95\% bootstrap CI) against mean tokens per issue. The dotted line is the mean of the L2R ranker over the five windows. Registry initialization is excluded.}
\label{fig:cost}
\end{figure}

\paragraph*{Token use} The coordinator configurations use 0.65M to 1.12M tokens per issue, compared with 0.32M for the Haiku RLM, about 1k for a plain LLM call, about 4k for the file stages of Agentless and CoSIL, 0.29M for RRL, and 1.37M for LocAgent (Figure~\ref{fig:cost}). Token use is not held fixed across the sequence, because it follows from what the agent does within the same step limit. The coordinator configurations use more tokens than the RLM, and part of the harness effect may come from the larger budget. Among the coordinator configurations, the budget differs little and does not explain the small differences between them.

\finding{Answer to RQ2. With the model held fixed, most of the gain comes from multi-turn exploration of the repository. Pooled F1 rises from 0.199 for the plain model to 0.598 with a REPL and to 0.677 when the same REPL runs in the coordinator's harness. Adding repository tools that limit their output changes it by only +0.019, which is not significant, and the subsystem registry, through its map or with consultation, does not help.}

\subsection{Guiding the Coordinator (RQ3)}
\label{sec:rq3}

\begin{table*}[t]
\centering
\caption{Guiding the coordinator (RQ3), pooled over all 105 issues under $G_{SRC}$, with three runs of the default system and one run of each variant. Consult. is the mean number of consultation calls per issue and Tokens the mean tokens per issue. $\Delta_{\mu}$ and $\Delta_{i}$ are the differences from the default system in pooled micro-F1 and in mean per-issue F1, with 95\% paired bootstrap intervals, W/L counts issues better and worse than the default, and $p$ is from the two-sided Wilcoxon signed-rank test. The lower rows combine the final outputs of the default system and of Agentless without changing the coordinator. $^\dagger$Answers with the Agentless files when Agentless and CoSIL return the same set (17\% of issue runs) and with the default system otherwise, saving 15\% of its tokens.}
\label{tab:guidance}
\small
\setlength{\tabcolsep}{3.5pt}
\begin{tabular}{lrrrrrccrr}
\toprule
Variant & F1 & P & R & Consult. & Tokens & $\Delta_{\mu}$ [95\% CI] & $\Delta_{i}$ [95\% CI] & W/L & $p$ \\
\midrule
Default (adaptive) & .664 & .655 & .674 & 0.72 & 1.12M & -- & -- & -- & -- \\
Nudged consultation & .676 & .686 & .667 & 1.40 & 1.39M & +0.012 [$-$0.026, +0.048] & +0.004 [$-$0.033, +0.037] & 38/36 & 0.56 \\
Seeded with Agentless & .671 & .651 & .691 & 0.46 & 1.27M & +0.006 [$-$0.033, +0.045] & $-$0.001 [$-$0.033, +0.031] & 38/35 & 0.90 \\
\midrule
Union with Agentless & .509 & .390 & .734 & & & & & & \\
Intersection with Agentless & .540 & .815 & .403 & & & & & & \\
Cascade$^\dagger$ & .607 & .588 & .629 & & & & & & \\
\bottomrule
\end{tabular}
\end{table*}

\paragraph*{Forced consultation} The nudged instruction made the coordinator consult the domain agents in 96\% of the issues, with 1.40 consultations per issue, compared with 49\% and 0.72 under the adaptive policy (Table~\ref{tab:guidance}). Pooled F1 changed by +0.012 [$-$0.026, +0.048] and per-issue F1 by +0.004 [$-$0.033, +0.037] ($p=0.562$), at higher token use. The point estimate favors forced consultation, but the difference is not significant, the interval includes zero, and the variant was run once. Our data therefore show no significant improvement from forcing consultation, and they do not exclude a small one, nor that another form of consultation could help.

\paragraph*{Candidates from a cheaper localizer}
\label{sec:seed}
With the files of Agentless in its prompt, the coordinator's pooled F1 changed by +0.006 [$-$0.033, +0.045] and its per-issue F1 by $-$0.001 [$-$0.033, +0.031] ($p=0.895$). This variant was run once, so its interval reflects the sampling of issues only. Combining the final outputs of the default system and Agentless does not help either (Table~\ref{tab:guidance}). The union raises recall but lowers precision, the intersection has the highest precision of all configurations but low recall, and the cascade saves tokens at a lower F1.

\finding{Answer to RQ3. Guidance did not help significantly. Neither forcing the coordinator to consult the domain agents on every issue nor giving it the candidate files of Agentless improved accuracy, and forced consultation (0.676) still scored below the coordinator alone (0.695). With one run per variant, a small gain from forced consultation cannot be ruled out.}

\subsection{Files Missed by All Methods (RQ4)}
\label{sec:rq4}

\begin{table*}[t]
\centering
\caption{Existing-file and new-file localization. Existing-file F1 is the micro-F1 of the predictions that exist at the base commit, scored against $G_{PR}^{exist}$. New recall is the share of files created by the resolving PRs whose exact path was predicted, separately for the 10 new source files (by the $G_{SRC}$ rule) and the 175 new tests, changelog fragments, and other files. The last three columns count predicted paths that neither exist at the base commit nor are created by the PR, per issue, split into proposed tests, changelogs, and other non-source files, and source paths, and give the share of all predicted paths that are such source paths.}
\label{tab:newfiles}
\small
\setlength{\tabcolsep}{3.5pt}
\resizebox{\textwidth}{!}{%
\begin{tabular}{lrrrrrrrrrr}
\toprule
& \multicolumn{5}{c}{Existing-file F1} & \multicolumn{2}{c}{New recall} & \multicolumn{3}{c}{Non-existent paths} \\
\cmidrule(lr){2-6}\cmidrule(lr){7-8}\cmidrule(lr){9-11}
Method & Ans.\ 2020 & Ans.\ 2025 & Ans.\ 2026 & OpenLib. & Qutebr. & Source & Other & Test etc./iss. & Source/iss. & Source share \\
\midrule
Plain LLM (Haiku) & .323 & .270 & .297 & .161 & .165 & 10\% & 1\% & 1.88 & 2.57 & 36\% \\
Plain LLM (Sonnet) & .368 & .284 & .374 & .264 & .425 & 17\% & 1\% & 0.71 & 0.95 & 20\% \\
Single-agent RLM (Haiku) & .463 & .542 & .571 & .571 & .686 & 17\% & 1\% & 0.30 & 0.19 & 3\% \\
Single-agent RLM (Sonnet) & .546 & .631 & .631 & .644 & .790 & 30\% & 2\% & 0.10 & 0.05 & 1\% \\
Coordinator, REPL only (Haiku) & .512 & .564 & .616 & .641 & .730 & 30\% & 1\% & 0.19 & 0.09 & 2\% \\
Coordinator only (Haiku) & .544 & .556 & .606 & .663 & .753 & 23\% & 1\% & 0.26 & 0.08 & 1\% \\
Coordinator + registry map (Haiku) & .563 & .529 & .535 & .633 & .720 & 20\% & 1\% & 0.19 & 0.08 & 1\% \\
Domain agents, adaptive (Haiku) & .516 & .538 & .561 & .629 & .730 & 23\% & 2\% & 0.38 & 0.12 & 2\% \\
Codex CLI (GPT-6.1-Terra, high) & .585 & .656 & .691 & .715 & .757 & 30\% & 1\% & 0.37 & 0.01 & 0\% \\
Reformulate-Retrieve-Localize & .288 & .212 & .284 & .415 & .269 & 10\% & 0\% & 0.00 & 0.01 & 0\% \\
Agentless (file stage) & .322 & .259 & .363 & .307 & .300 & 0\% & 0\% & 0.00 & 0.00 & 0\% \\
CoSIL (file stage) & .316 & .273 & .354 & .305 & .277 & 0\% & 0\% & 0.00 & 0.00 & 0\% \\
LocAgent & .398 & .358 & .479 & .498 & .407 & 0\% & 0\% & 0.00 & 0.00 & 0\% \\
\bottomrule
\end{tabular}}
\end{table*}

Newly created files are rarely predicted. The resolving PRs create 10 new source files and 175 new tests, changelog fragments, and other files. The best method predicts 30\% of the new source files and 2\% of the others by their exact path (Table~\ref{tab:newfiles}). Agentless, CoSIL, LocAgent, and the non-LLM rankers select only from files that exist, so they cannot predict a new file. RRL ends with an agent that writes its list freely, and in all three runs it named the file created by the one Open Library fix that only adds files, which gives it one of the 10 new source files. The task prompt asks for test and documentation files, which invites the agents to propose new test modules and changelog fragments. Their names follow project conventions that the agents rarely match.

Predicted paths that neither exist at the base commit nor are created by the fix fall into two groups. Proposed new tests, changelog fragments, and other non-source files make up 5\% of the paths predicted by the coordinator. Source paths that do not exist, which are the paths recalled from other versions of a project or invented, make up 36\% of the paths predicted by the Haiku plain LLM and 20\% for the Sonnet plain LLM, but 1\% for the coordinator and 3\% for the RLM. Reading the repository therefore largely removes this error.

Two windows contain documentation co-changes, Ansible 2020 and qutebrowser (Table~\ref{tab:docs}). On Ansible, documentation recall is much lower than code recall for every method, for example 12\% for the domain agents and 23\% for Codex. On qutebrowser, the agentic systems recover most documentation files, for example 61\% for the domain agents and 91\% for Codex. Nine of the 11 qutebrowser documentation changes are to \texttt{doc/help/settings.asciidoc}, a reference page generated from the settings definitions in \texttt{qutebrowser/config/configdata.yml}, which always changes together with that file. The Ansible documentation changes are prose, namely porting guides, user guides, and shared snippets spread over 15 different files, and the issue texts do not mention them.

\begin{table}[t]
\centering
\caption{Documentation co-change. Recall of code and of documentation files in $G_{SRC}$ for Ansible 2020 (50 code, 16 documentation files) and qutebrowser (59 code, 11 documentation files).}
\label{tab:docs}
\scriptsize
\resizebox{\columnwidth}{!}{%
\begin{tabular}{lrrrr}
\toprule
& \multicolumn{2}{c}{Ansible 2020} & \multicolumn{2}{c}{qutebrowser} \\
\cmidrule(lr){2-3}\cmidrule(lr){4-5}
Method & Code & Docs & Code & Docs \\
\midrule
Plain LLM (Haiku) & 41\% & 2\% & 11\% & 15\% \\
Plain LLM (Sonnet) & 61\% & 4\% & 36\% & 45\% \\
Single-agent RLM (Haiku) & 62\% & 17\% & 73\% & 52\% \\
Single-agent RLM (Sonnet) & 60\% & 25\% & 81\% & 55\% \\
Coordinator, REPL only (Haiku) & 67\% & 23\% & 78\% & 58\% \\
Coordinator only (Haiku) & 71\% & 29\% & 82\% & 85\% \\
Coordinator + registry map (Haiku) & 73\% & 29\% & 79\% & 55\% \\
Domain agents, adaptive (Haiku) & 71\% & 12\% & 83\% & 61\% \\
Codex CLI (GPT-6.1-Terra, high) & 73\% & 23\% & 69\% & 91\% \\
Reformulate-Retrieve-Localize & 62\% & 0\% & 55\% & 0\% \\
Agentless (file stage) & 64\% & 0\% & 53\% & 0\% \\
CoSIL (file stage) & 66\% & 0\% & 56\% & 0\% \\
LocAgent & 78\% & 0\% & 75\% & 0\% \\
\bottomrule
\end{tabular}}
\end{table}

\finding{Answer to RQ4. New files are rarely predicted, and most non-existent paths that the agents predict are proposed tests and changelogs rather than recalled source paths. Documentation is recovered when it is structurally coupled to a source file and largely missed when it is narrative documentation that the issue does not mention.}

\subsection{Robustness Across Repositories, Time, and Issue Texts (RQ5)}
\label{sec:rq5}

\begin{figure*}[t]
\centering
\includegraphics[width=0.92\textwidth]{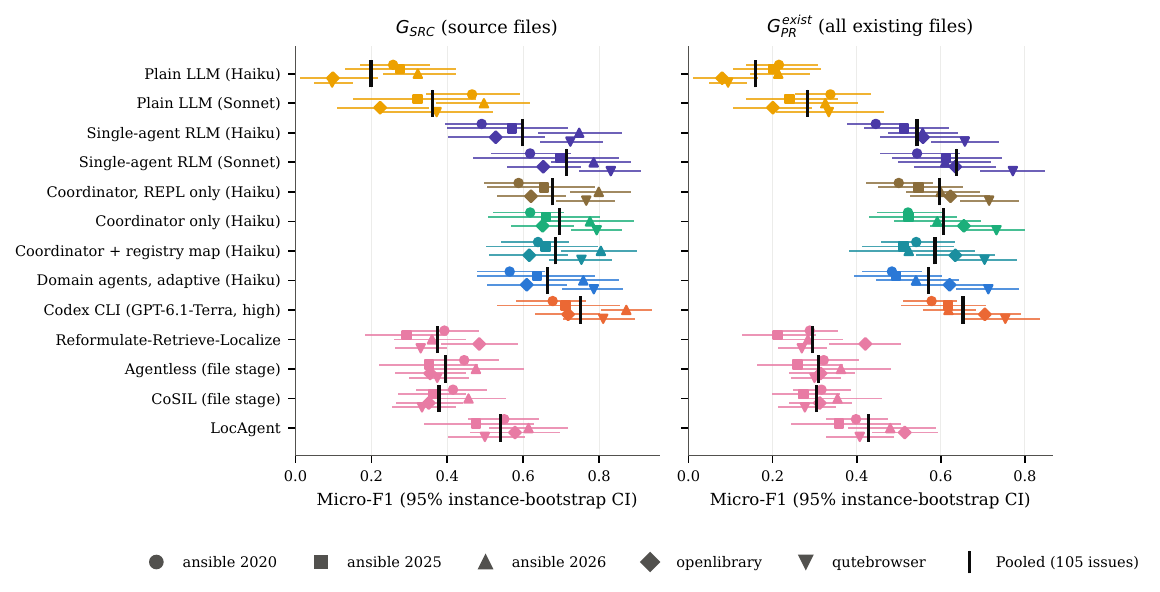}
\caption{Micro-F1 per window with 95\% issue-bootstrap intervals, and pooled micro-F1 over all 105 issues (vertical bar), under $\Gsrc$ (left) and $\Gex$ (right).}
\label{fig:forest}
\end{figure*}

Figure~\ref{fig:forest} shows the results per window with bootstrap intervals. The main conclusion holds everywhere. In every repository and window, the exploring configurations are well above the plain models, and the coordinator configurations are close to each other. The comparisons with LocAgent and Codex are less robust. On the 67 issues whose texts come from SWE-bench Pro, the coordinator reaches 0.689, the domain agents 0.653, the RLM 0.575, LocAgent at the held-out cutoff 0.584, and Codex 0.736 (Table~\ref{tab:subsets}). Here the held-out cutoff of LocAgent is chosen on the other SWE-bench Pro windows only. On this subset the coordinator remains significantly better than the RLM ($p_{Holm}=0.002$), whereas the differences of the domain-agent system from LocAgent at the held-out cutoff and from Codex keep their sign but are not significant with 67 issues. Per window, LocAgent at the held-out cutoff scores higher than the domain agents on Ansible 2020 and Open Library, so its overall deficit comes from the other windows. The overlap analysis of Section~\ref{sec:benchmark} concerns the identifiers that the generated texts name. It cannot show whether the diff informed these texts in other ways. The plain LLMs vary most across windows and do much better on Ansible than on the other repositories. Their higher scores on the Ansible windows of 2025 and 2026, whose issue texts were written for this study and some of whose issues fall within the models' training period, should not be read as evidence about memorized fixes. Absolute F1 differs between windows, for example because the Ansible 2026 issues have smaller gold sets, so comparisons should be made within a window.

\finding{Answer to RQ5. The advantage of the exploring configurations over the plain LLMs holds on all three repositories, in all five windows, and on the 67 issues without generated texts. The comparisons with LocAgent and Codex hold only in aggregate, as detailed above. The size of the plain LLMs' advantage on Ansible points to familiarity with a widely used project, but our windows do not separate familiarity from memorized fixes.}

%% ---------------------------------------------------------------------------
\section{Discussion~\texorpdfstring{\protect\aicite}{}}
\label{sec:discussion}

\subsection{Interpretation}
Our results point to one conclusion. What makes a small model a good localizer is the freedom to explore the repository over many turns, not the structure imposed on that exploration. A study of tool-using agents in customer-service workflows reached a similar conclusion, finding that added orchestration did not improve task success on average and helped mainly where the agent lacked information~\cite{lim2026declarative}. The model that reads only the issue reaches 0.199 pooled F1, and the same model with a REPL in an agent harness reaches 0.677. The structure we added on top changed little. Without output-limited tools the model wrote equivalent code in the REPL, the registry lowered pooled F1 slightly, and guidance brought no significant gain, although a modest gain from forced consultation remains compatible with our single run. Once the RLM sees the output of each action, it is a strong baseline, and the differences between the coordinator configurations fall within the uncertainty of the data.

The comparison with the published localizers is consistent with this view. Agentless and CoSIL choose files mainly from the repository structure, RRL from BM25 results and file beginnings, and LocAgent through graph search, each with a limited number of interactions. All of the multi-turn configurations we tested are above them with the same model, including LocAgent scored at its best held-out cutoff. Larger models still help, and the Sonnet RLM and Codex are the most accurate configurations.

\subsection{Implications}
\paragraph*{For tool builders} For multi-file localization, give the model multi-turn access to the repository in an agent loop with native tool calling, and invest in the model rather than in elaborate exploration structure. Output-limited tools and registries may still help with longer contexts or larger repositories than ours, where unlimited reads would exhaust the context window, and the agents in our study kept their reads small on their own. Predicted paths should be checked for existence, because plain LLMs produce many non-existent source paths.

\paragraph*{For researchers and benchmark designers} Agentic baselines must be checked for how they actually run. A runner that lets the model answer before it has seen any output, crashes that are scored as empty predictions, and checkouts that contain later history can each create or hide differences between configurations. Studies should report how runs ended, the access that agents had to later history and to the network, and the tool logs that show what they did. They should report the gold-set definition, separate new-file and existing-file localization, keep pooled micro-F1 apart from per-issue averages, and report paired intervals for differences. Evaluations should include more than one project and strong lexical baselines tuned without test data, which reach F1 0.448 to 0.500 on two of our three repositories.

\paragraph*{For practitioners} New test files and narrative documentation are the files that automated localization misses most often. Generated reference documentation is found reliably because it is coupled to a single source file.

\subsection{Cost}
At 0.65M to 1.12M tokens per issue, localization with a coordinator agent is expensive, and the RLM reaches most of its accuracy with 0.32M. Most of these tokens are re-sent conversation history that is read from the prompt cache, which makes them much cheaper than their count suggests. When cost is the main constraint, a single agent exploring with a REPL is the best trade-off in our data, and Agentless or a lexical ranker remains a reasonable first-stage filter, particularly on repositories where issue texts share vocabulary with the code. Al Awad and Ivanov~\cite{alawad2026costeffective} similarly find that cheaper explorers keep most of the localization quality of expensive ones.

%% ---------------------------------------------------------------------------
\section{Threats to Validity~\texorpdfstring{\protect\aicite}{}}
\label{sec:ttv}

\paragraph*{Construct validity} A gold set derived from a merged PR treats every changed file as required, although some changes may be incidental. We address this by reporting four gold definitions and showing where the conclusions depend on them. All-gold is a strict measure for issues with many gold files, so we report it together with micro-F1. One Ansible 2020 SWE-bench Pro instance (\texttt{502270c}) has a base commit dated 2021 while its issue date is in 2020, and we keep it as distributed. The 2025 and 2026 Ansible issue texts were written by a Codex worker that could read the resolving PR, including its diff (Section~\ref{sec:benchmark}). They name no gold file, and they name the identifiers changed by the fix about as often as the SWE-bench Pro texts do, but their uniform and well-structured style differs from that of user-written issues and may favor all LLM methods. We therefore report every result also on the 67 issues without them (Table~\ref{tab:subsets}). The SWE-bench Pro issue texts of the added repositories name some gold files (10 of 150 $\Gsrc$ files by full path), which helps every method that reads the text, including the lexical rankers. Under $\Gsrc$ we remove predicted tests, changelogs, and metadata before scoring. This favors methods that list such files, and without it the domain agents, which often list the tests a fix would update, would score lower. We apply the filter to every method and report $\Gex$ without it. The published localizers index only Python files, so we also report a Python-only evaluation, and they return ranked lists, so we also score them at a held-out cutoff.

\paragraph*{Internal validity} All agentic configurations reported here were run in the same way, with per-issue single-commit checkouts, the access controls of Section~\ref{sec:protocol}, a 600\,s timeout for the coordinator and domain-agent configurations and for Codex, three runs per window (one for the Sonnet RLM and the two RQ3 variants, because of cost), and a registry built afresh in every run. Only the coordinator with and without the repository tools differ in a single component (Section~\ref{sec:rq2}). Registry construction is itself random, with 8 to 45 agents per registry, which adds variation between runs that the run-level intervals include. Agents may run code in their REPL, and the guard refuses only the actions listed in Section~\ref{sec:protocol}. Every tool call of the domain agents is logged with its output and was audited (Section~\ref{sec:protocol}). Models may have seen older fixes during training, and RQ5 reports the results on issues filed after the training cutoffs.

The published localizers were designed and tuned with larger models (for example GPT-4o or Claude 3.5 Sonnet) and are run here with Haiku 4.5 to keep the model fixed. They may perform better with the models of their original evaluations. We changed their code only to load our instances, to add the model, and, for LocAgent, to repair malformed tool arguments, and we reimplemented RRL from its description because no code is available. All changes are included in the replication package.

\paragraph*{External validity} The 105 issues come from three Python repositories with different architectures. The results may differ for other languages, for much larger repositories, or for projects without clear subsystem boundaries, where subsystem decomposition may help less. Our repositories are of moderate size, and structured exploration may matter more in much larger ones, where unrestricted reads would fill the context window sooner. We use one model family for the agentic systems and one commercial agent. Localization is evaluated in isolation, and its effect on end-to-end repair is not measured.

\paragraph*{Conclusion validity} With 105 issues, small effects can still go undetected, and comparisons within a single window have low power. We use non-parametric paired tests, correct for 21 planned comparisons, and report effect sizes, issue-level bootstrap intervals, and paired intervals for the differences. A difference that is not significant is not evidence that two configurations are equivalent, and we state the interval that the data allow. The guidance variants of RQ3 were tested with one instruction each, and other forms of guidance may behave differently.

%% ---------------------------------------------------------------------------
\section{Conclusion~\texorpdfstring{\protect\aicite}{}}
\label{sec:conclusion}
We asked which parts of the exploration structure of LLM agents matter for multi-file change localization, on 105 issues from three Python repositories and five time windows, with every agent working in a checkout of the issue's base commit. With a small model held fixed, the answer is the multi-turn exploration loop itself. Letting the model explore the repository through a REPL, and running it in an agent harness with native tool calling, accounts for most of the gain. The structure we added on top, repository tools that limit their output and a registry of subsystem-scoped domain agents, brought no measurable improvement, and the registry lowered pooled F1 slightly. Forcing consultation, run once, showed no significant gain, though a modest one remains possible. The resulting small-model agents exceed four published localizers run with the same model, while larger models remain more accurate. Newly created files and narrative documentation remain largely unsolved by every method, whereas documentation generated from code is found reliably. Future work should test whether structured exploration helps in larger repositories and with longer contexts, cover other languages, and measure the effect of localization quality on end-to-end repair.

\section*{Acknowledgment}
During the preparation of this work the authors used Claude Code (Anthropic) to re-score the experimental runs, implement the classical baselines and the reimplementation of RRL, adapt and run the published localizers, correct the experiment harness, add its access controls, and run and audit the experiments of this revision. The original version of this article was written by the authors. The authors used Claude Code (Anthropic)~\cite{claudecode} to generate initial drafts of its revisions, in the Abstract, in all sections, which cite it in their headings, and in the response to the reviewers. The authors reviewed and revised those passages and take responsibility for the final text. The issue texts of the 2025 and 2026 Ansible windows were written with Codex (OpenAI) from the resolving pull requests and their discussions, as described in Section~\ref{sec:benchmark}. After using these tools, the authors reviewed and edited all of the content and take full responsibility for the content of the published article.

The replication package contains the benchmark windows (instance identifiers, base commits, issue texts, and all gold sets), every per-issue prediction and token log, including those of the published localizers, the per-issue tool logs of all agentic runs, the domain-agent registries, the audit of the first-submission runs, the overlap analysis of the issue texts, the prompts, our changes to the published localizers, the RRL reimplementation, and the scripts that regenerate all tables, figures, and numbers in the text. It is available at \url{https://github.com/CodeAssureLabs/subsystem-agents-localization} (release used for this article: commit \texttt{d1704ac}). The agent implementation used for all LLM runs, including the selection and issue-generation code of the 2025 and 2026 windows, is available at \url{https://github.com/CodeAssureLabs/swe-agent-spawning} (commit \texttt{9b18b88}).


\begin{IEEEbiography}[{\includegraphics[width=1in,height=1.25in,clip,keepaspectratio]{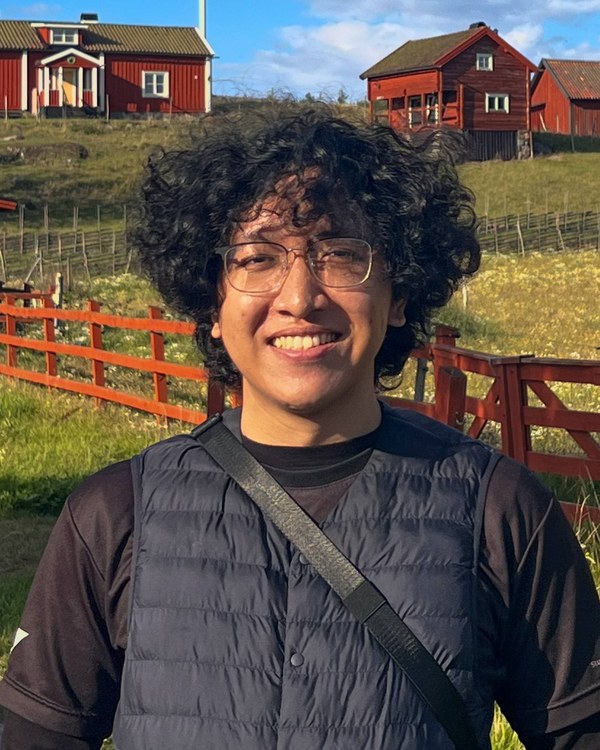}}]{Akeela Darryl Fattha}
graduated summa cum laude in Information Systems from Singapore Management University, Singapore, in 2026, and is the founder of Homegrid Labs, a property platform.
\end{IEEEbiography}

\begin{IEEEbiography}[{\includegraphics[width=1in,height=1.25in,clip,keepaspectratio]{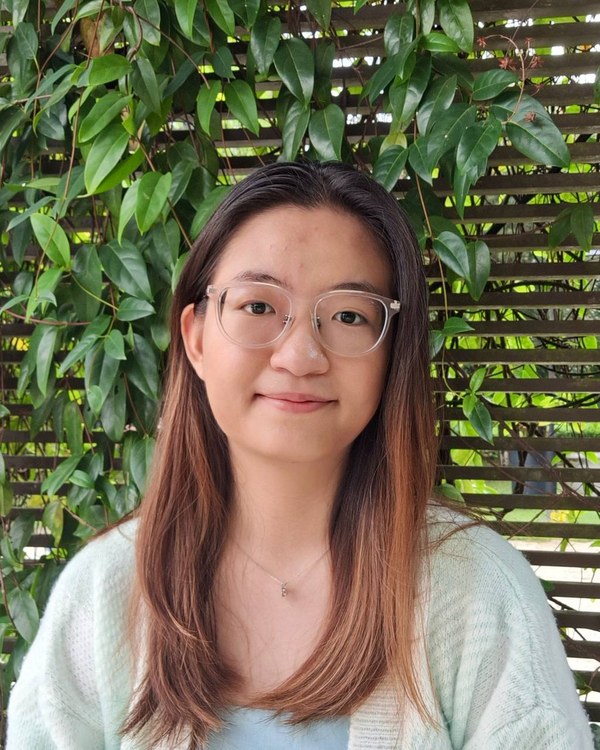}}]{Kia Ying Chua}
is pursuing a Bachelor of Science in Computer Science at Singapore Management University, Singapore. Her interests include AI, machine learning, and their applications.
\end{IEEEbiography}

\begin{IEEEbiography}[{\includegraphics[width=1in,height=1.25in,clip,keepaspectratio]{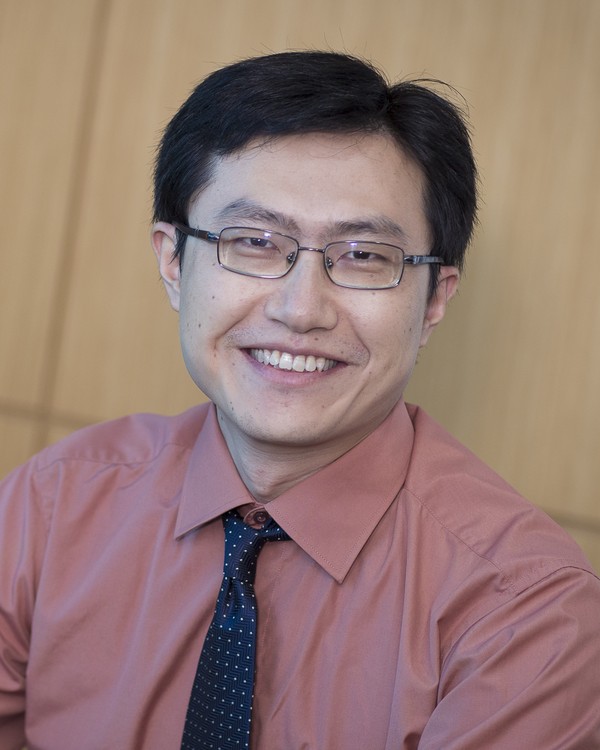}}]{Lingxiao Jiang}
is a Professor of Computer Science in the School of Computing and Information Systems at Singapore Management University, Singapore, and the Director of the Centre for Research for Intelligent Software Engineering (RISE). Jiang's research interests include software engineering, software security, and data mining for software.
\end{IEEEbiography}

\begin{IEEEbiography}[{\includegraphics[width=1in,height=1.25in,clip,keepaspectratio]{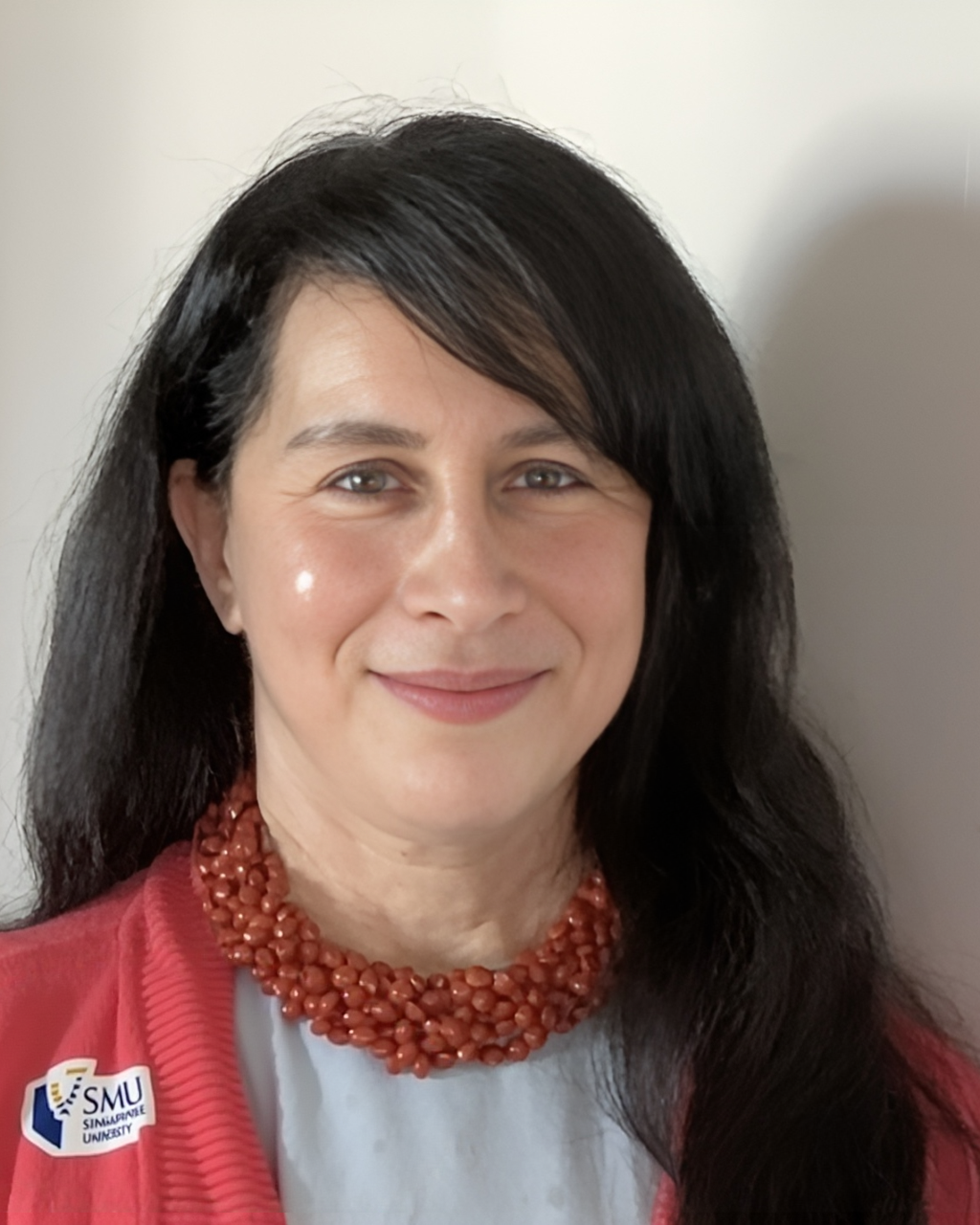}}]{Laura Wynter,}
PhD, is Associate Professor of Computer Science (Practice) in the School of Computing and
Information Systems at Singapore Management University, Singapore. Previously she led the IBM
Research lab in Singapore. Her research interests are agentic AI and applications of AI.
\end{IEEEbiography}

\EOD

\end{document}